\documentclass[11pt]{article}
\usepackage[a4paper,margin=2.5cm]{geometry} \usepackage[T1]{fontenc} \usepackage[utf8]{inputenc} \usepackage{authblk} \usepackage{graphicx} \usepackage{amsmath,amssymb} \usepackage{booktabs} \usepackage{textcomp} \graphicspath{{./figures/}} 
\usepackage[percent]{overpic}
\usepackage[title]{appendix}%
\usepackage{xcolor}%
\usepackage{float}%
\usepackage{color}
\usepackage{tikz}
\usepackage{cite}
\usepackage{pgfplots}
\pgfplotsset{compat=1.18}
\usepackage[mathlines]{lineno}
\usepackage{hyperref}
\hypersetup{
    colorlinks,
    citecolor=blue,
    filecolor=blue,
    linkcolor=blue,
    urlcolor=blue
}

\pgfplotsset{compat=1.18}
\pgfplotsset{
  natureplot/.style={
    width=0.99\linewidth,
    ymajorgrids=true,
    grid style={gray!25},
    axis line style={black},
    tick style={black},
    tick align=outside,
    line width=0.9pt,
  }
}

\usepackage{booktabs}
\usepackage{longtable}
\usepackage{array}
\usepackage{graphicx}
\usepackage{pgfplots}
\pgfplotsset{compat=1.18}
\usepackage{caption}
\usepackage{float}
\usepackage{siunitx}
\usepackage{threeparttable}
\usepackage{rotating}
\renewcommand{\arraystretch}{1.12}

\newcommand{\nuc}[2]{\ensuremath{^{#1}\mathrm{#2}}}
\newcommand{\qion}[3]{\ensuremath{^{#1}\mathrm{#2}^{#3+}}}

\newcommand{\sigfitsys}{\ensuremath{\sigma_{\mathrm{fit+sys}}}}

\begin{document}
\title{Precision masses of neutron-rich platinum and gold nuclei reveal enhanced $N=126$ shell strength below doubly-magic $^{208}$Pb}

\author[1,2,*]{David Freire-Fern\'andez} 
\author[3,1,*]{Rui-Jiu Chen}
\author[4]{Usama Ahmed}
\author[1]{Helena M. Albers} 
\author[1,5]{Jelena Bardak}
\author[1]{Carsten Brandau}
\author[4,1]{Jeroen P. Bormans} 
\author[6]{R. Burcu Cakirli}
\author[7]{Rikel Chakma}
\author[8,24]{Maeve Cockshutt}
\author[8,24]{Iris Dillmann}
\author[1,25]{Dmytro Dmytriiev}
\author[9]{Siddharth Doshi}
\author[1,2]{Carlo Forconi}
\author[1,10]{Oliver Forstner}
\author[3]{Wenwen Ge}
\author[1]{Jan Glorius}
\author[1]{Magdalena G{\'o}rska}
\author[8]{Chris J. Griffin}
\author[1]{Alexandre Gumberidze}
\author[1]{Regina Hess}
\author[1,26]{Pierre-Michel Hillenbrand}
\author[1]{Nicolas J. Hubbard}
\author[11]{Calum Jones}
\author[12]{Beatriz Jurado}
\author[13]{Kanika}
\author[7]{Filip G. Kondev}
\author[14]{Gregor Kosir}
\author[15,*]{Wolfram Korten}
\author[1]{Christophor Kozhuharov}
\author[4,16,1]{Johan Emil Larsson}
\author[12]{Guy Leckenby}
\author[3]{Hongfu Li}
\author[17]{Menglan Liu}
\author[1]{Sergey Litvinov}
\author[1,2,*]{Yuri A. Litvinov}
\author[3]{Zhong Liu}
\author[1]{Bernd Lorentz}
\author[4]{Hannes Mayr}
\author[1,2]{Esther B. Menz}
\author[18]{Tetsuaki Moriguchi}
\author[4]{Clemens M. Nickel}
\author[11]{Zachary Nunns}
\author[19]{Fatma Cagla Ozturk}
\author[1]{Nikolaos Petridis}
\author[11]{Zsolt Podolyak}
\author[20,1]{Shahab Sanjari}
\author[11]{Ragandeep Singh Sidhu}
\author[1,$\dagger$]{Markus Steck} 
\author[1,10,21]{Thomas St{\"o}hlker}
\author[14]{Jelena Vesic}
\author[3]{Meng Wang}
\author[3]{Qian Wang}
\author[11]{Philip M. Walker}
\author[1]{Helmut Weick}
\author[1,2]{Michael Weinert}
\author[2]{Kathrin Wimmer}
\author[12,26,1]{Boguslaw W{\l}och}
\author[3]{Xing Xu}
\author[22]{Takayuki Yamaguchi}
\author[3]{Xinliang Yan}
\author[3]{Yue Yu}
\author[23]{Cenxi Yuan}
\author[3]{Min Zhang}
\author[3]{Yuhu Zhang}
\author[3]{Xu Zhou}

\affil[1]{GSI Helmholtzzentrum f\"ur Schwerionenforschung GmbH, 64291 Darmstadt, Germany}
\affil[2]{Institut f\"ur Kernphysik, Universit{\"a}t zu Köln, 50937 Cologne, Germany}
\affil[3]{State Key Laboratory of Heavy Ion Science and Technology, Institute of Modern Physics, Chinese Academy of Sciences, 730000 Lanzhou, China}
\affil[4]{Institut f{\"u}r Kernphysik, Technische Universit\"at Darmstadt, 64289 Darmstadt, Germany}
\affil[5]{Department of Physics, University of Novi Sad, 21000 Novi Sad, Serbia}
\affil[6]{Max-Planck Institut f\"ur Kernphysik, 69117 Heidelberg, Germany}
\affil[7]{Physics Division, Argonne National Laboratory, Lemont, Illinois 60439, USA}
\affil[8]{TRIUMF, Vancouver, BC V6T 2A3, Canada}
\affil[9]{School of Architecture, Technology and Engineering, University of Brighton, Brighton, BN2 4GJ, UK}
\affil[10]{Helmholtz-Institut Jena, 07743 Jena, Germany}
\affil[11]{School of Mathematics and Physics, University of Surrey, Guildford, GU2 7XH, UK}
\affil[12]{Universit\'e de Bordeaux, CNRS, LP2i Bordeaux, 33170 Gradignan, France}
\affil[13]{Imperial College London, London, SW7 2AZ, United Kingdom}
\affil[14]{Jo\v{z}ef Stefan Institute, 1000 Ljubljana, Slovenia}
\affil[15]{IRFU, CEA, Universit\'e Paris-Saclay, 91191 Gif-sur-Yvette, France}
\affil[16]{Helmholtz Forschungsakademie Hessen für FAIR (HFHF), GSI Helmholtzzentrum für Schwerionenforschung GmbH, 64291 Darmstadt, Germany}
\affil[17]{Institute for Basic Science, 34126 Daejeon, Republic of Korea}
\affil[18]{Institute of Pure and Applied Sciences, University of Tsukuba, Tsukuba 305-8571, Japan}
\affil[19]{Department of Physics, Istanbul University, 34134 Istanbul, Turkey}
\affil[20]{Waldorf School Frankfurt am Main, 60433 Frankfurt am Main, Germany}

\affil[21]{Institute of Optics and Quantum Electronics, Friedrich Schiller University Jena, 07743 Jena, Germany}

\affil[22]{Department of Physics, Saitama University, Saitama 338-8570, Japan}
\affil[23]{Sino-French Institute of Nuclear Engineering and Technology, Sun Yat-sen University, Zhuhai, Guangdong 519082, China}

\affil[24]{Department of Physics and Astronomy, University of Victoria, Victoria, BC V8P 5C2, Canada}
\affil[25]{Deutsches Elektronen Synchrotron DESY, 15738 Zeuthen, Germany}
\affil[26]{Facility for Antiproton and Ion Research in Europe GmbH, 64291
Darmstadt, Germany}

\affil[*]{Corresponding authors: D.FreireFernandez@gsi.de, ChenRJ13@impcas.ac.cn,
W.Korten@cea.fr,
Y.Litvinov@gsi.de}
\affil[$\dagger$]{Deceased.}

\maketitle

\begin{abstract}
The heaviest stable nuclei in the universe owe their existence to quantum shell 
structure, the grouping of protons and neutrons into discrete energy levels 
separated by gaps\,\cite{Otsuka-2020}. The largest known neutron shell gap in stable nuclei, at 
$N=126$, stabilizes doubly-magic $^{208}$Pb and is responsible for the 
characteristic abundance peak of heavy elements near gold and platinum produced 
by the rapid neutron-capture process (r-process)\,\cite{Cowan-2021}. Whether this shell gap 
persists as protons are removed from lead is a question central to both nuclear 
structure\,\cite{Yuan-2022} and the modeling of heavy-element synthesis\,\cite{Li-2026}, yet it has remained 
unanswered due to the extraordinary difficulty of producing the relevant 
neutron-rich nuclei. Direct experimental knowledge in this region was essentially 
absent. Here we report the first precision mass measurements of 
$^{203,204}$Pt and $^{204,205,206}$Au, performed at GSI using a novel 
combination of Schottky and isochronous mass spectrometry in a heavy-ion storage 
ring.
The $N=126$ isotones $^{204}$Pt and $^{205}$Au are more strongly bound than the extrapolated trend of the previously known mass surface\,\cite{AME2020} by 403 and 464~keV, respectively, revealing an unexpectedly enhanced $N=126$ shell strength below doubly-magic $^{208}$Pb. 
Furthermore, the proton-neutron interaction 
strength\,\cite{Zhang-1989} exhibits a hitherto unobserved bifurcation at $N=126$ as protons are 
removed from $^{208}$Pb. Our results redefine the nuclear mass surface in the 
neutron-rich heavy-element region and provide direct experimental benchmarks for 
theoretical models whose extrapolations toward more exotic nuclei are essential 
for r-process nucleosynthesis calculations.
\end{abstract}

\section{Shell evolution below \(^{208}\)Pb} \label{sec:shell_evolution} 

Shell closures are among the clearest signatures of quantum many-body structure in atomic nuclei. They produce enhanced binding and characteristic discontinuities in one- and two-nucleon separation energies, giving rise to the traditional magic numbers \(2,8,20,28,50,82\) and \(126\). These numbers were explained by the nuclear shell model as the consequence of large gaps between single-particle orbitals \cite{MMG,Haxel}. Because nuclear binding energies are obtained directly from measured masses, precision mass spectrometry provides one of the most sensitive probes of shell evolution far from stability \cite{Blaum-2006,Yamaguchi-2021}. 

The magic numbers were established near stability, but radioactive-beam experiments have shown that shell structure can change substantially in exotic nuclei\,\cite{Kanungo-2009, Otsuka-2020, Ye-2025}. Classical shell closures such as \(N=8\) and \(N=20\) weaken in neutron-rich systems, whereas new closures emerge at \(N=14,16,32\) and \(34\), and a proton shell closure at \(Z=14\) has recently been identified in \(^{22}\)Si \cite{Xing-2025}. These observations demonstrate that shell gaps are not immutable constants, but depend on the balance of single-particle energies, correlations, deformation and proton--neutron interactions. 

Shell structure in neutron-rich nuclei is also central to the rapid neutron-capture process, \(r\)-process, which is responsible for producing about half of the elements heavier than iron. This gives rise to element-abundance peaks near \(A\sim80\), 130 and 195 associated with the \(N=50\), 82 and 126 shell closures, respectively \cite{B2FH,Cowan-2021}. The \(N=126\) closure is particularly important for the formation of the third abundance peak, yet it remains the least experimentally constrained of the major neutron shell closures in the heavy neutron-rich region.

\begin{figure}[] 
\centering 
\includegraphics[width=1\textwidth]{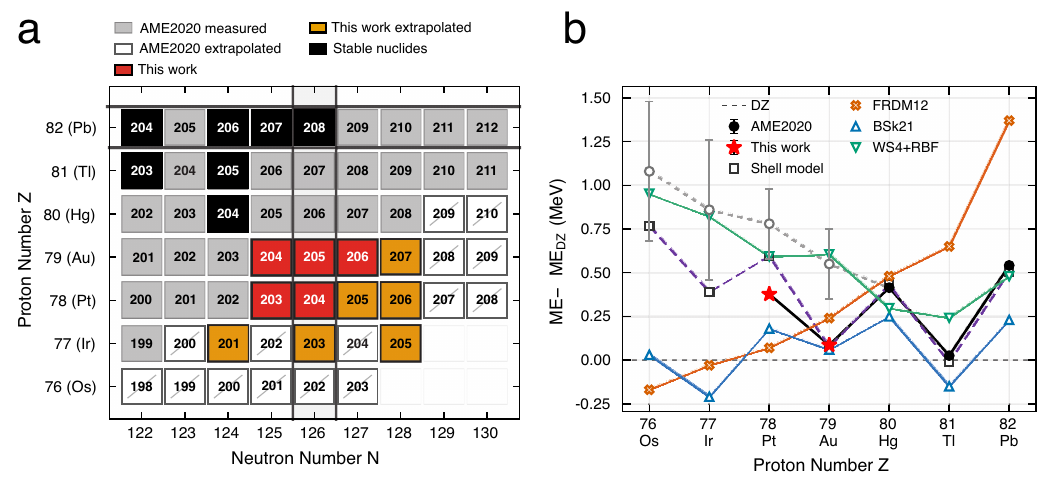} 
\caption{\textbf{Nuclear masses below doubly-magic $^{208}$Pb.} \textbf{(a)} A part of nuclear chart around \(N=126\) illustrating nuclear masses from the Atomic Mass Evaluation (AME2020)\,\cite{AME2020}. Black, gray and white colors indicate known masses for stable and radioactive nuclei, and extrapolated masses, respectively. The five masses measured in this work are shown in red. 
Updated AME values are drawn in orange (see text). 
\textbf{(b)} Mass-excess residuals of \(N=126\) isotones relative to the
Duflo--Zuker (DZ) mass model, \(\Delta ME=ME-ME_{\mathrm{DZ}}\). Black filled circles denote AME2020 values based on experimental input, 
whereas grey open circles denote AME2020 extrapolated values\,\cite{AME2020}; red stars denote the new \(^{204}\mathrm{Pt}\) and \(^{205}\mathrm{Au}\) masses.
Representative global and shell-model calculations are also shown. The new masses replace extrapolated anchors below mercury and show additional binding at \(N=126\).} 
\label{fig:masses}
\end{figure}

The scarcity of data below \(^{208}\)Pb reflects the difficulty of producing and identifying these nuclei.
Below doubly magic \(^{208}\)Pb, directly measured masses of \(N=126\) isotones had been available only for the neighboring nuclei \(^{207}\)Tl and \(^{206}\)Hg, see Fig.\,\ref{fig:masses} (a).
The $^{206}$Hg mass was determined directly only recently\,\cite{Goodacre-2021}, replacing the value based on the \(^{210}\)Pb\((\alpha)^{206}\)Hg \(\alpha\)-decay link  measured more than six decades ago \cite{206Hg}. As a result, the evolution of the \(N=126\) shell closure toward lower proton numbers has so far been inferred mostly from 
theoretical models. However, it is well known that global mass models diverge rapidly in regions lacking experimental data\,\cite{Sobichewski-2018,Buskirk-2024}.

Known masses and their extrapolations had suggested a reduction of the \(N=126\) shell strength below \(^{208}\mathrm{Pb}\), in analogy with the shell gap reduction observed in the isotones above \(^{208}\mathrm{Pb}\). However, \(r\)-process simulations based on such a weakened shell closure reproduce the third abundance peak only under highly neutron-rich conditions and modified \(\beta\)-decay rates \cite{Li-2026}. Direct mass measurements below mercury are therefore needed both to test the local shell evolution and to benchmark the models used for extrapolations toward the astrophysical \(r\)-process path. 


Here, we report the first mass measurements of \(^{203,204}\)Pt$_{125, 126}$ and \(^{204,205,206}\)Au$_{125, 126, 127}$. 
The nuclei \(^{204}\)Pt and \(^{205}\)Au double the number of directly measured \(N=126\) isotones below lead, extending experimental knowledge two proton numbers beyond mercury, see Fig.\,\ref{fig:masses}\,(a). The neighboring \(N=125\) and \(N=127\) masses determine the local $_{78}$Pt and $_{79}$Au mass-surface slopes across the shell closure. Together, these data test the persistence of the \(N=126\) shell below $_{82}$Pb and provide new benchmarks for models in this region of the mass surface.

A comparison of the known \(N=126\) masses with representative calculations is shown in Fig.~\ref{fig:masses}\,(b). All values are displayed as mass-excess residuals relative to the Duflo--Zuker (DZ) mass model\,\cite{DZ}. The models include the Finite-Range Droplet Model (FRDM12)\,\cite{FRDM}, the Weizs\"acker--Skyrme mass model with radial-basis-function correction (WS4+RBF)\,\cite{WS4}, 
the Hartree-Fock-Bogoliubov model with the BSk21 functional (BSk21)\,\cite{BSk21}, and shell-model calculations constrained near \(^{208}\mathrm{Pb}\)\,\cite{Yuan-2022}. The spread among model predictions grows rapidly toward lower proton numbers, illustrating the need for direct experimental anchors below mercury.

\section{Experimental details and data analysis} \label{sec:schottky_ims} 

\begin{figure}[p] 
\centering 
\includegraphics[width=0.9\textwidth]{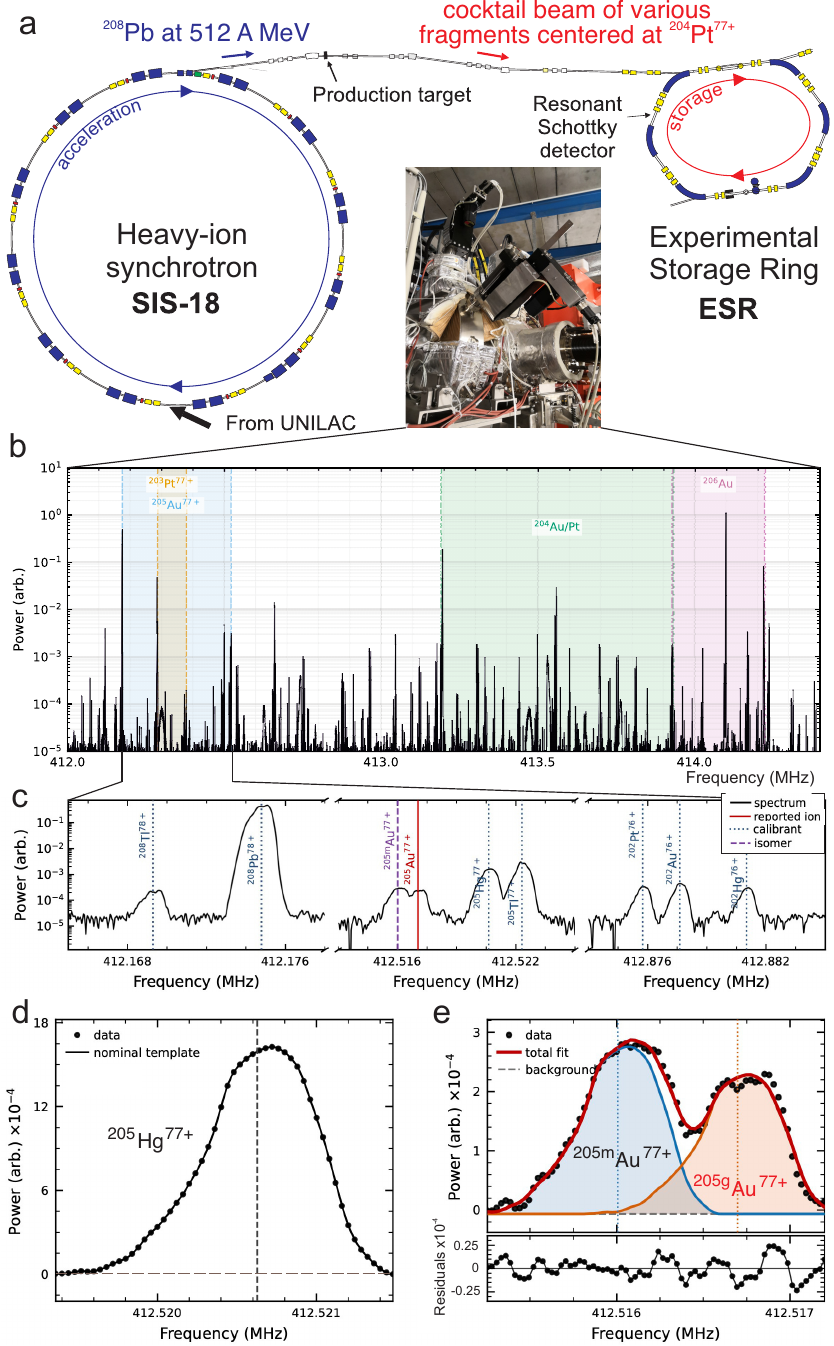} 
\caption*{\textit{(Figure \ref{fig:experiment} caption on next page)}} \label{fig:experiment}
\end{figure} 

\begin{figure}[] 
\addtocounter{figure}{-1} 
\caption{\textbf{Production, storage and non-destructive detection of neutron-rich ions.} 
\textbf{(a)} A \(^{208}\)Pb beam at \(512\,A\)MeV was fragmented in a \(^{9}\)Be target and the resulting cocktail beam was injected into the experimental storage ring ESR.  The transport line and the ESR were tuned centering on \(^{204}\mathrm{Pt}^{77+}\) ions. 
Other nuclei with known and unknown masses 
within the ion-optical acceptance of the machine were transmitted and stored as well. 
The insert shows the cavity-based Schottky detector resonant at \(\sim413.5\)\,MHz, which was utilized to measure the revolution frequencies of the stored ions. \textbf{(b)} The averaged broadband Schottky frequency spectrum (note the logarithmic scale). 
\textbf{(c)} Example of local frequency regions selected for mass calibration (note the interrupted frequency scale). 
The regions shown correspond to harmonic number \(h=214\) (see text) and  
are for the \(^{205}\mathrm{Au}^{77+}\) ions.
Vertical markers denote the centroids employed in the mass calibration, where the reference masses (black) are used to determine the unknown mass of interest (red).
\textbf{(d)} An isolated peak of \(^{205}\mathrm{Hg}^{77+}\) ions utilized to define the peak shape template in this local frequency region (note the linear scale). 
\textbf{(e)} Decomposition of the overlapping peaks of \(^{205m}\mathrm{Au}^{77+}\) (isomeric state, orange) and \(^{205g}\mathrm{Au}^{77+}\) (ground state, blue) ions. The total fit (red) describes well the measured data (black points) as confirmed by the residuals plot. 
} \label{fig:experiment} 
\end{figure}

The experiment was performed at GSI Helmholtzzentrum f\"ur Schwerionenforschung in Darmstadt, Germany (GSI). The relevant part of the facility is shown schematically in Fig.~\ref{fig:experiment} (a). A \(^{208}\)Pb primary beam was accelerated to \(512\,A\)MeV by the UNILAC linear accelerator and the SIS-18 heavy-ion synchrotron, rebunched and extracted every \(\sim4\) s in short pulses of approximately \(0.5~\mathrm{\mu}\mathrm{s}\). The extracted beams impinged on a \(1.8~\mathrm{g\,cm^{-2}}\) \(^{9}\)Be production target. Projectile fragmentation was the dominant nuclear reaction producing a cocktail of heavy neutron-rich fragments. 
After the target, the fragments emerged as highly charged ions with no or only a few bound electrons\,\cite{scheidenberger1998}. The ion-optical system was centred on hydrogen-like \(^{204}\mathrm{Pt}^{77+}\), thereby defining the central magnetic rigidity ($B\rho$) of the transport line, \[ B\rho=\frac{p}{q}=\frac{mv\gamma}{q}, \] where \(p\), \(m\), \(q\), \(v\) and \(\gamma\) are the ion momentum, mass, charge state, velocity and relativistic Lorentz factor, respectively.
The fragments produced in the nuclear reactions had large momentum spreads. 
Consequently, multiple nuclear species within the magnetic-rigidity acceptance were simultaneously transported and injected into the Experimental Storage Ring (ESR)\,\cite{Franzke-1987}.

Production of exotic nuclei at relativistic energies is well studied at GSI\,\cite{geissel1995}. Removing protons, while retaining nearly all neutrons from \(^{208}\)Pb, produced neutron-rich Pt and Au isotopes near \(N=126\), which is a rare cold fragmentation reaction\,\cite{Benlliure-1999}. Access to \(N=127\) in \(^{206}\)Au requires charge-exchange channels\,\cite{Rodriguez-2020}. 
The production rates of all nuclear species,
their atomic charge state distributions, and transmission efficiencies were simulated with the transport code LISE++\,\cite{tarasov2008_LISE, Tarasov_2016}.

In the ESR, for two stored ions with different mass-to-charge ratios, the relative revolution-frequency difference is\,\cite{Franzke-2008} 
\begin{equation} 
\frac{\Delta f}{f} = -\alpha_p\frac{\Delta(m/q)}{m/q} + \left(1-\frac{\gamma^2}{\gamma_t^2}\right)\frac{\Delta v}{v}, \label{eq:freq_moq} 
\end{equation} 

where \(\alpha_p=1/\gamma_t^2\) is the momentum-compaction factor and \(\gamma_t\) is the transition Lorentz factor of the ring\,\cite{Franzke-2008}. The ESR was operated in an isochronous ion-optical mode\,\cite{Steck-2020}, see Methods. For ions fulfilling \(\gamma\simeq\gamma_t=1.4\), the velocity-dependent term vanishes to first order and the revolution frequency becomes a
measure for \(m/q\). The primary beam energy was chosen such that the energy of \(^{204}\mathrm{Pt}^{77+}\) ions after the production target, taking into account nuclear reaction kinematics and energy loss processes, matched the isochronicity condition. Small inhomogeneities of magnetic fields of the ESR magnets result in \(\gamma_t\) variations across the ESR aperture\,\cite{Hausmann-2000}. To maximize the mass resolving power, the ESR acceptance was reduced with mechanical scrapers, while the width and shape of the stored-ion frequency distributions were monitored online. 

The combined Schottky+Isochronous Mass Spectrometry (S+IMS) technique\,\cite{Tu-2018, Fernandez-2024}, see Methods, has been utilized to measure revolution frequencies of stored ions non-destructively.
For this purpose, a 410\,MHz resonant Schottky cavity was used\,\cite{Sanjari2020}. 
In short, every particle circulated in the ESR with a revolution frequency of about 2\,MHz, thereby exciting the cavity with a pulse-like current, known as Schottky noise. The resulting signal is a superposition of the Schottky and thermal noise.
The tunable resonance frequency was set to \(413.5\) MHz, corresponding to the overlapping \(213^{\rm th}\) and \(214^{\rm th}\) harmonics of the revolution frequencies of the stored ions. 
The signal from the cavity was amplified and sampled. The subsequent Fast Fourier Transform (FFT) was used to create revolution frequency spectra, see Fig.\,\ref{fig:experiment} (b). According to Eq.\ref{eq:freq_moq}, the revolution frequencies contain information on the particle mass-to-charge ratio. The areas under the frequency peaks reflect the corresponding numbers of stored ions\,\cite{Litvinov-2011, Bosch-2013}.  
The parameters of the FFT were optimized offline.
The technique is capable of measuring the revolution frequency of a single stored ion within a few milliseconds\,\cite{Fernandez-2024}. 
This work represents the first application of the S+IMS technique to broadband mass measurements.

A total of 11,452 injections of fresh ions into the ESR were processed and the revolution frequency spectra were analyzed. Frequency drifts in different spectra, that are common for all peaks, were corrected. These drifts were due to instabilities of the ESR magnets and the activation of the septum magnet at each injection. The aligned spectra were summed together. The resulting spectrum was used for the identification of frequency peaks.
This procedure is well established in isochronous mass spectrometry\,\cite{Tu-2011b, Xing-2019}, see Methods.
An additional advantage of the S+IMS is that the same spectrum is measured at several harmonics thereby providing an independent identification cross-check.

The slits selected a narrow range from the broad momentum distributions of the injected particles. Hence, it can safely be assumed that all orbits allowed by the slits were populated with the same probability. The measured frequency distribution is non-Gaussian and reflects the remaining nonlinearities of $\gamma_t$. The shape of the peaks alters slowly as a function of frequency as a consequence of moving away from the isochronous region. However, within a small range of frequencies the peak shape is preserved. The regions used to determine the masses reported here are indicated in Fig.\,\ref{fig:experiment} (b) and one of them is zoomed in Fig.\,\ref{fig:experiment} (c). All the other regions used in this work are shown in Extended Data Fig.\,\ref{fig:extended_local_spectra}.

In addition to the ions of interest, there were multiple ion species present in the spectra which have well-known masses. In particular, less-exotic isobars in the same atomic charge state lie in immediate proximity to the ions of interest, see Fig.\,\ref{fig:experiment} (b-c). The nuclides with known masses were used for calibrating the frequency scale, see Methods.

A representative example of the peak shape is shown in Fig.\,\ref{fig:experiment} (d) for the case of $^{205}$Hg$^{77+}$. This peak was used to construct a template to fit all peak shapes within the region.
The quality of the method is illustrated in Fig.\,\ref{fig:experiment} (e) where two components of the overlapping peaks of the ground and isomeric states of $^{205}$Au$^{77+}$ are clearly decomposed.
The known mass-to-charge ratios, taken from the latest Atomic Mass Evaluation (AME2020)\,\cite{AME2020}, were fitted with low-order polynomials to obtain the unknown masses, see Methods. The reliability of the approach was tested by re-determining the known masses.
More details on the template construction, ion identification, reference-ion selection, fitting procedure and error propagation can be found in Methods and Supplementary Materials.

\section{Mass surface across \(N=126\)} \label{sec:masses} 

We report ground-state masses of neutron-rich platinum and gold nuclei spanning the \(N=126\) shell closure. The measured set comprises the $N=125$ isotones, \(^{203}\)Pt and \(^{204}\)Au, the $N=126$ isotones, \(^{204}\)Pt and \(^{205}\)Au, and the $N=127$ gold isotope, \(^{206}\)Au. The data therefore constrain the local platinum and gold mass surface on both sides of \(N=126\).

The results are summarized in Table~\ref{tab:masses}. The quoted uncertainties, denoted here by \(\sigma_{\mathrm{fit+sys}}\), include statistical uncertainties of the peak and calibration fitting (\(\sigma_{\mathrm{fit}}\)). The systematic uncertainty (\(\sigma_{\mathrm{sys}}\)) was estimated with a standard procedure adopted in storage ring mass spectrometry\,\cite{Litvinov-2005b}: The mass of each reference ion was re-determined assuming it is unknown, see Methods, and the obtained values were compared to the literature ones. 
The systematical uncertainty (\(\sigma_{sys}\)) was calculated by forcing the corresponding reduced \(\chi^2_\nu=1\).

Preliminary mass excess values of ME$(\rm ^{204}Au)=-20648(24)$\,keV and  ME$(\rm ^{205}Au)=-19067(31)$\,keV were reported in a doctoral thesis\,\cite{Amanbayev-2023}. These masses were measured by a different technique applying multi-reflection time-of-flight spectrometer at the ion-catcher setup of GSI\,\cite{Plass-2013, Dickel-2015}. Our values agree, respectively, within $1.25\sigma$ and $1.01\sigma$. Hence, in the case of $^{205}$Au, both measurements show a large deviation to the AME2020 extrapolation.

Several isomeric states were observed in the spectra, as illustrated by the $^{205m}$Au-$^{205g}$Au doublet in Fig.\,\ref{fig:experiment} (e). The $^{205m}$Au state was known previously from spectroscopic studies~\cite{PODOLYAK2009116}, but the present measurement provides the direct determination of its mass. The excitation energy obtained from our ground-state mass, \(E_x(^{205m}\mathrm{Au})=930(14)~\mathrm{keV}\), agrees within $1.54\sigma$ with the value \(E_x=907(5)\)\,keV reported in the NUBASE2020 evaluation\,\cite{Kondev-2021}, providing an additional consistency check of the peak assignment and centroid extraction routines. 
The aforementioned thesis work\,\cite{Amanbayev-2023} also reports the excitation energy of the $^{205m}$Au isomer extracted from the measured masses of \(E_x=953(53)\)\,keV, which is in excellent agreement ($-0.43\sigma$) with our value. 
Observed isomeric states were included in the spectral decomposition procedure to avoid biasing the ground-state centroids. The present work focuses on the ground-state masses listed in Table~\ref{tab:table1}; extracted data on isomeric states will be discussed elsewhere.
Details on all calibrations and the corresponding uncertainties are provided in the Supplementary Materials.

\begin{table}[t] 
\centering 
\caption{\textbf{Ground-state masses across \(N=126\).} Mass excesses from this work (\(\mathrm{ME}_{\mathrm{GSI}}\)) are compared to literature values taken from AME2020 (\(\mathrm{ME}_{\mathrm{AME2020}}\))\,\cite{AME2020}. 
The symbol \(\#\) denotes extrapolated values.
The uncertainty of \(\Delta \mathrm{ME}\) is dominated by the uncertainties of extrapolated AME2020 values.} \label{tab:masses} \renewcommand{\arraystretch}{1.15} 
\begin{tabular}{lcccc} \hline 
Atom & \(N\) & \(\mathrm{ME}_{\mathrm{GSI}}\) (keV) & \(\mathrm{ME}_{\mathrm{AME2020}}\) (keV) & \(\Delta \mathrm{ME}\) (keV) \\ \hline 
\(^{203}\mathrm{Pt}\) & 125 & \(-19742(37)\)  & \(-19510(200)\#\) & \(-232(203)\#\) \\ 
\(^{204}\mathrm{Pt}\) & 126 & \(-18023(7)\)  & \(-17620(200)\#\) & \(-403(200)\#\) \\ 
\(^{204}\mathrm{Au}\) & 125 & \(-20618(2)\)  & \(-20390(200)\#\) & \(-228(200)\#\) \\ 
\(^{205}\mathrm{Au}\) & 126 & \(-19034(10)\)  & \(-18570(200)\#\) & \(-464(200)\#\) \\ 
\(^{206}\mathrm{Au}\) & 127 & \(-14322(21)\)  & \(-14190(300)\#\) & \(-133(301)\#\) \\ 
\hline \end{tabular} 
\label{tab:table1}
\end{table}

\section{Increased shell strength below Pb}
\label{sec:shell_interpretation}

The new masses for all five nuclei shown in Table~\ref{tab:table1} indicate that these nuclei are more strongly bound than the respective extrapolations in the AME2020. 
In particular, the stronger binding of $N=126$ Au and Pt isotones alters the previously deduced weakening trend of this neutron shell\,\cite{Li-2026}. 

\begin{figure}[t] 
\centering 
\includegraphics[width=1\textwidth]{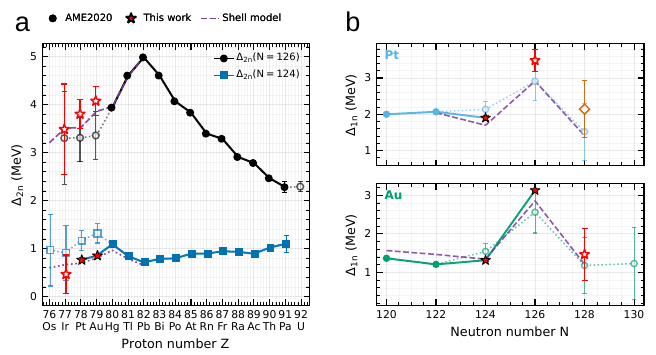} 
\caption{\textbf{Neutron shell-gap indicators.} \textbf{(a)} The empirical shell gap \(\Delta_{2n}(Z,N=126)\) and the shifted gap \(\Delta_{2n}(Z, N=124)\) values are shown as a function of proton number. Filled symbols denote values based on experimentally known masses in AME2020. 
Open symbols show the data containing extrapolated data in AME2020. 
Red symbols denote data obtained in this work. Purely experimental values are shown in full symbols and data containing the updated AME extrapolations, see Methods, are shown with open symbols.
The shell model calculations\,\cite{Yuan-2022} are added for reference.
\textbf{(b)} One-neutron shell-gap \(\Delta_{1n}\) for platinum (top) and gold (bottom) isotopes as a function of neutron number.
The notation is the same as in panel (a). 
The red diamond symbol indicates the value extracted solely from updated AME extrapolations.
} \label{fig:deltan} 
\end{figure}

Various finite-difference mass filters are implemented to isolate nuclear structure properties, see Methods. To quantify the shell strength, the empirical shell gap \(\Delta_{2n}\) is typically used, which is the difference of two-neutron separation energies.
The required \(N=128\) masses for \(^{206}\mathrm{Pt}\) and \(^{207}\mathrm{Au}\) are not yet experimentally known and, hence, the extrapolated values need to be taken. Because the new masses differ substantially from the AME2020 extrapolations, a local update of the AME was carried out to achieve the most reliable predictions of the required masses, see Methods.
The employed AME database contains all experimental data available including the masses reported here. The obtained values for Ir, Pt and Au isotopes are given in Methods and will be published in the forthcoming issue of the AME, AME2026.
The resulting \(N=126\) shell-gap indicator is shown in Fig.~\ref{fig:deltan} (a). The new masses raise the Pt and Au shell-gap values by \(0.496\) and \(0.717\,\mathrm{MeV}\), respectively, relative to the AME2020 baseline, reversing the smooth weakening trend previously inferred from extrapolated masses.
The effect of a stronger binding of $N=126$ isotones is clearly seen in the change of the downward trend of $\Delta_{2n}$ as compared to AME2020. The asymmetry in behavior of the shell gap values for $N=126$ isotones below and above doubly-magic $^{208}$Pb is striking.

In regions at the edge of the known mass surface, the availability of experimental masses to extract \(\Delta_{2n}\) is often very limited. 
In such cases, other mass filters can be applied to estimate the robustness of shell closures. 
We considered two complementary finite-difference observables that test the local mass surface curvature
with reduced dependence on unmeasured neighbours.

First, a so-termed shifted gap $\Delta_{2n}(N=124)$ can be considered to obtain qualitative information on the evolution of $\Delta_{2n}(N=126)$. It has been shown for both, proton and neutron shell closures\,\cite{Manea-2023}, that the shifted gap follows the trend of the empirical shell gap with an opposite sign. This implies that the increased $N=126$ shell strength in Au shall be accompanied by a decreased shifted gap at $N=124$. The shifted gap is illustrated in the same figure with the empirical shell gap data (Fig.~\ref{fig:deltan} (a)), where the abrupt change in Au is clearly observed.

Second, one-neutron shell-gap indicator \(\Delta_{1n}\), defined as a finite difference of one-neutron separation energies, 
provides a
direct local test of the separation-energy discontinuity across a given neutron number.
The \(\Delta_{1n}\) values are plotted in Fig.~\ref{fig:deltan} (b), where the enhanced shell strength in Au and Pt is clearly seen in comparison to the AME2020 baseline. 

The stronger $N=126$ shell has important consequences for the $r$-process. The $r$-process simulations assuming the decreasing shell trend given by the AME2020, see Fig.\,\ref{fig:deltan} (a), found it challenging to reproduce the narrow $A=195$ peak of solar elemental abundances\,\cite{Li-2026}. Our results alter these conclusions. Quantifying the precise impact will require dedicated network calculations with the updated mass surface, which we encourage.

The even $N=126$ $^{206}$Hg isotone was investigated with \((d,p)\) transfer reactions to measure binding energies of several single-particle neutron orbitals in the $N+1$ $^{207}$Hg isotope lying beyond the closed shell\,\cite{Tang-2020}. The results obtained, combined with data on lead and polonium isotones, allowed the establishment of trends for these orbitals toward lighter isotones, thereby predicting that \(N=126\) nuclei become unbound with respect to neutron emission ($S_n=0$) at around $Z\sim64$. As a consequence, the $r$-process path near the $N=126$ waiting point is constrained by nuclear structure in a way that strongly depends on the single-particle level ordering, creating a bottleneck more significant than the one at $N=82$\,\cite{Mumpower-2016}. The baseline for these studies is given by the respective ground state masses. In this context, our results lead to a shallower trend, hence, keeping the rest of the discussion in \cite{Tang-2020} unaltered, the limit of the nuclear existence would be shifted toward even lower-Z nuclei.

Large-scale shell-model calculations with a dedicated Hamiltonian were reported recently\,\cite{Yuan-2022}, in which the entire region ``south'' of $^{208}$Pb was examined, including the $_{74}$W-$_{82}$Pb $N=126$ isotones. One key result of that study is that the $N=126$ shell gap is predicted to be robust from $Z=82$ down to $Z=68$ with only a minor reduction. Our results confirm this conclusion, although our data indicate a trend toward a stronger shell closure. A small, but physically important, 0.1\,MeV modification to the proton-proton two-body matrix elements was introduced specifically to improve the description of binding energies in the Au, Pt, and Ir isotopes~\cite{Yuan-2022}. 
This shows that even state-of-the-art shell models require empirical tuning in this region. 
The corresponding shell model results are included in Figs.\,\ref{fig:masses} (b) and \ref{fig:deltan} (a).
Our masses now provide the first direct experimental constraints to guide future refinements of this interaction. 

In this work we updated the aforementioned shell-model calculations. 
The Yuan22 interaction, see Methods, provides the best overall description of the new Pt and Au ground-state binding energies. While the modified interaction improves the description of selected excited states in \(^{204}\)Pt, in particular through changes in the \((\pi s_{1/2}d_{5/2})\) monopole matrix elements, it worsens the calculated ground-state binding energy of \(^{204}\)Pt. 

Within the present uncertainties, and despite including one extrapolated mass in each $\Delta_{2n}(N=126)$ calculation for \(Z=78\) and \(Z=79\), the direct Pt and Au masses support a sizeable \(N=126\) shell effect at both platinum and gold as seen in Fig.~\ref{fig:deltan} (a, b). The additional binding at \(N=126\) cannot by itself identify the microscopic origin of the shell stabilization. It may reflect changes in single-particle spacings, quadrupole correlations, proton-neutron monopole terms or a combination of these effects \cite{FP-1978}. Furthermore, the measured masses replace extrapolated anchors in the region where model predictions diverge, and they show that the \(N=126\) shell effect remains robust at least down to platinum.

\begin{figure}[t] 
\centering 
\includegraphics[width=0.9\textwidth]{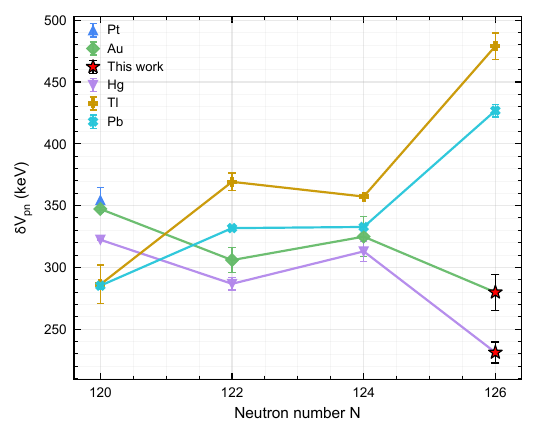} 
\caption{\textbf{Proton--neutron interactions near \(N=126\).} Average proton--neutron interaction strengths \(\delta V_{pn}\) are shown for Pt, Au, Hg, Tl and Pb nuclei with \(N\le126\). 
Red stars mark the values directly modified by the new masses: \(^{205}\mathrm{Au}\), which uses the new \(^{205}\mathrm{Au}\) and \(^{204}\mathrm{Pt}\) masses, and \(^{206}\mathrm{Hg}\), which uses the new \(^{204}\mathrm{Pt}\) mass. All other input masses are taken from AME2020\,\cite{AME2020}.
The new points reveal a bifurcation between the Au--Hg and Tl--Pb trends at \(N=126\), suggesting a local change in the proton--neutron interactions below mercury.} 
\label{fig:dvpn} 
\end{figure}

We also extracted the average proton-neutron interaction, \(\delta V_{pn}\), defined in the Methods, from the updated mass surface. Only experimental values, from this work and AME2020, were used. The resulting values are displayed in Fig.~\ref{fig:dvpn}. In particular, the values derived from the new platinum and gold masses reveal a bifurcation between the $_{79}$Au–$_{80}$Hg and $_{81}$Tl–$_{82}$Pb nuclei at \(N=126\). Observing such an effect for nuclei filling in the $N=82-126$ shell is an unexpected result and suggests that the proton–-neutron interaction evolves significantly as different orbitals are emptied just below the closed shells at $^{208}$Pb. To give a microscopic interpretation based on experimental data, we would require yet experimentally unavailable \(\delta V_{pn}\) data points for lighter isotones. Although such measurements are highly complex, the future data will elucidate whether the observed bifurcation is a part of a broader hole--hole proton--neutron correlation pattern below the \(N=126\) shell closure. In particular, direct masses of the next \(N=126\) isotones, \(^{203}\mathrm{Ir}\) and \(^{202}\mathrm{Os}\), together with their neighboring \(N=125\) and \(N=127\) masses, will be decisive for testing whether the observed trend persists below platinum.

The new \(^{204}\mathrm{Pt}\) and \(^{205}\mathrm{Au}\) masses increase the binding relative to the AME2020 extrapolations at \(N=126\) by $>400$\,keV. This additional binding strengthens the empirical shell-gap systematics below doubly-magic \(^{208}\mathrm{Pb}\), challenges extrapolations that predicted a smoother weakening of the shell closure, and provides direct benchmarks for local microscopic calculations and global mass models in this heavy neutron-rich region of the nuclear chart. Yet unknown masses of lower-\(Z\) \(N=126\) isotones and their neighbours are needed to conclude whether the persistent shell effect and the observed pattern of the proton--neutron interaction continue below platinum.



\newpage
\section{Methods}
\label{sec:methods}

\subsection{Mass filters}
Atomic masses can be straightforwardly converted to nuclear binding energies. The masses of bound electrons and their total electron binding energy need to be taken into account. For a nucleus with proton number \(Z\), neutron number \(N\), and atomic mass \(M_{\rm at}(Z,N)\), the nuclear binding energy is given as
\begin{equation}
B(Z,N)=Zm_pc^2+Nm_nc^2-M_{\rm nuc}(Z,N)c^2, 
\end{equation} 
where \(M_{\rm nuc}(Z,N)c^2=M_{\rm at}(Z,N)c^2-Zm_ec^2+E_{\rm bind}(Z)\). Here \(m_p\), \(m_n\) and \(m_e\) are the proton, neutron and electron rest masses, respectively, and \(E_{\rm bind}(Z)\) is the total electronic binding energy of the neutral atom\,\cite{Rodrigues-2004}. 

For the highly charged ions used in the mass calibration, the corresponding charge-state-dependent electron binding energies\,\cite{Rodrigues-2004} were included explicitly in the ionic mass-to-charge ratios. 

The one- and two-neutron separation energies are defined as 
\begin{equation} 
S_n(Z,N)=B(Z,N) - B(Z,N-1),
\end{equation} 
and 
\begin{equation} 
S_{2n}(Z,N)=B(Z,N) - B(Z,N-2).
\end{equation} 

The empirical neutron shell-gap indicator is conventionally given as a difference of two-neutron separation energies \begin{equation} 
\label{eqn:D2n}
\Delta_{2n}(Z,N)=S_{2n}(Z,N)-S_{2n}(Z,N+2). 
\end{equation} 

The one-neutron shell-gap indicator is defined as a finite difference of one-neutron separation energies
\begin{equation} 
\Delta_{1n}(Z,N)=S_{n}(Z,N)-S_{n}(Z,N+1). 
\end{equation} 

An additional indicator that can be employed to probe the shell effect is the single-particle neutron energy splitting at the Fermi level, defined in\,\cite{Buskirk-2024} as
\begin{equation} 
\Delta e_{n}(Z,N)=S_{n}(Z,N)-S_{n}(Z,N+2),
\end{equation} 
which is less practical in our context since it requires the masses of $N=128$ isotones.

The average proton--neutron interaction strength, $\delta V_{pn}$ \cite{Zhang-1989,Cakirli-2005,Brenner-2006}, can be extracted from double differences of binding-energies separately for even--even, even--odd, and odd--even 
systems, respectively 
\begin{align} 
\delta V_{pn}^{ee}(Z,N) &= \frac{1}{4} \left[ (B({Z,N})-B({Z,N-2}))-(B({Z-2,N})-B({Z-2,N-2})) \right], \\ 
\delta V_{pn}^{eo}(Z,N) &= \frac{1}{2} \left[ (B({Z,N})-B({Z,N-1}))-(B({Z-2,N})-B({Z-2,N-1})) \right], \\ 
\delta V_{pn}^{oe}(Z,N) &= \frac{1}{2} \left[ (B({Z,N})-B({Z,N-2}))-(B({Z-1,N})-B({Z-1,N-2})) \right]. 
\end{align} 

All finite-difference observables used here were obtained with standard Gaussian uncertainty propagation. The quoted uncertainties include the new mass uncertainties and, where applicable, AME2020 uncertainties for neighboring or extrapolated masses.

\subsection{Combined Schottky+Isochronous Mass Spectrometry}
Mass measurements were performed at the Experimental Storage Ring (ESR) at GSI, Darmstadt, using a novel combination of Schottky and Isochronous mass spectrometry (S+IMS)\,\cite{Tu-2018, Fernandez-2024}. As described by Eq.\,\ref{eq:freq_moq}, the revolution frequency of a stored ion depends on both its mass-to-charge ratio $m/q$ and its velocity. The key challenge is that the large velocity spread of secondary reaction products -- typically $\Delta v/v \sim 10^{-3}$ -- broadens the frequency distribution and limits mass resolving power unless the velocity dependence is eliminated.

In conventional Schottky mass spectrometry (SMS)\,\cite{Radon-1997}, the velocity spread is reduced to $\Delta v/v \sim 10^{-7}$ by electron cooling\,\cite{Steck-2004}, after which the revolution frequency becomes an accurate measure of $m/q$ alone. However, electron cooling takes several seconds, making SMS unsuitable for the short-lived nuclei.

In isochronous mass spectrometry (IMS)\,\cite{Hausmann-2001, Stadlmann-2004}, no cooling is required. Instead, the ring is tuned to a special ion-optical mode in which the Lorentz factor of the stored ions equals the transition energy parameter of the ring, $\gamma  = \gamma_t$\,\cite{Franzke-2008}. Under this condition, faster ions travel on longer orbits and slower ions on shorter ones, such that all ions with the same $m/q$ ratio circulate with the same revolution frequency regardless of their velocity. Frequencies are measured using a time-of-flight detector equipped with a thin carbon foil\,\cite{Trotscher-1992, Mei-2010}. However, due to energy loss in the foil, this technique is quasi-destructive -- ions are lost within a few hundred revolutions\,\cite{Zhang-2016} -- precluding lifetime measurements alongside mass determination.

A non-destructive Schottky detection\,\cite{schottky} can be a way to depart from such ``quasi-destructive'' measurements. 
However, the first generation of capacitive detectors required tens of seconds for detecting a single ion\,\cite{Litvinov-2004a}.
The newly developed resonant cavity-based detectors boosted the sensitivity\,\cite{Nolden-2011, Sanjari-2020}, thereby enabling their application to studying short-lived nuclei.
In the S+IMS technique employed here, such Schottky diagnostic is applied in the isochronous mode. 
Non-destructive detection enables the mass and lifetime of each stored ion to be determined in a single measurement -- a capability not available with conventional IMS. The isochronous tuning of the ring minimizes the velocity dependence of the frequency, allowing the Schottky signal to serve as a direct mass measurement without the need for cooling.

The S+IMS technique was pioneered at the ESR of GSI\,\cite{Walker-2013} and later employed at the experimental cooler-storage ring CSRe at HIRFL center in Lanzhou\,\cite{Tu-2018}. The development of highly-sensitive Schottky cavities allowed access to short-lived nuclei\,\cite{Nolden-2011, Sanjari-2020}. The first application was the measurement of the isolated nuclear 2-photon decay in $^{72}$Ge~\cite{Fernandez-2024}. Here we report the first use of the S+IMS technique for broadband mass measurements.

\subsection{Resonant cavity Schottky detector}
Revolution frequencies of stored ions were measured using a resonant cavity Schottky pickup installed in the ESR arc section\,\cite{Sanjari-2020}. A photograph of the detector is depicted in the insert of Fig.\,\ref{fig:experiment} (a). The detector is a cavity resonator operating at 410\,MHz, designed to couple to the longitudinal Schottky noise signal induced by the periodic passage of stored ions. Each stored ion, circulating with a characteristic revolution frequency, generates a weak periodic electromagnetic (EM) signal in the cavity. The cavity resonates at harmonics of the revolution frequency, amplifying the signal and enabling detection of single ions.
The detector features a quality factor $Q\sim 3000$ and a simulated shunt impedance of 473\,k$\Omega$. Two independent tuning mechanisms are implemented. The resonance frequency is adjusted by motorized copper plungers inserted into the cavity volume, which modify the effective cavity volume and hence the resonant frequency. 
The $Q$ value -- and therefore the sensitivity -- is varied by motorized insertion of a dissipative blade at the cavity rim where the magnetic field of the fundamental mode is maximum, reducing the $Q$ without significantly affecting the characteristic shunt impedance.

The signal from the cavity is amplified and Fourier analyzed, producing a frequency spectrum in which each stored ion species appears as a narrow peak at its characteristic revolution frequency and its harmonics. The non-destructive nature of the detection -- the ions are not disturbed by the measurement -- allows continuous monitoring of each ion's frequency over its entire storage lifetime, enabling simultaneous determination of mass and nuclear lifetime within a single measurement.

In addition to the 410\,MHz cavity, an older less-sensitive 245\,MHz cavity was used as a back-up detector\,\cite{Nolden-2011}.

\subsection{Data processing}

Two commercial spectrum analyzers were used to sample amplified signals from 245\,MHz\,\cite{Nolden-2011} and 410\,MHz\,\cite{Schottky1, Sanjari-2020} cavities producing complex data consisting of in-phase and quadrature (IQ) components. They were triggered by an injection event provided by the GSI accelerator control system. In addition, a continuous acquisition system NTCAP\,\cite{NTCAP} recorded the IQ data at a sampling rate of \(10~\mathrm{MS\,s^{-1}}\). 
In total, 11,452 injections were recorded.

Data from the 410\,MHz cavity were used in the present analysis. 
Each IQ record was transformed into a spectrogram utilizing Fast Fourier Transform (FFT). Non-overlapping 25-ms frames and the Blackman--Harris windowing function were used. With the \(10~\mathrm{MS\,s^{-1}}\) sampling rate, each FFT frame contained \(250{,}000\) complex samples, corresponding to a frequency-bin width of \(40~\mathrm{Hz}\). 
The first 0.5\,s in every injection were excluded from the analysis to reduce sensitivity to very short-lived species and to remove electronic distortion due to activated injection septum magnet.
The final revolution frequency spectra were in the time interval \(0.5~\mathrm{s}\leq t\leq3.75~\mathrm{s}\). This retained 130 frames per injection. 
Altogether, \(1{,}488{,}760\) individual spectra were produced.

Small variations of magnetic fields of the ESR cause temporal drifts of revolution frequencies common for all ion species. The high intensity peaks were used to correct for these drifts. Afterwards the spectra were averaged.




A robust penalized least-squares estimator \cite{baseline} removed the smooth detector baseline. The baseline was constrained by a second-derivative smoothness penalty and iteratively reweighted so that narrow positive Schottky peaks did not bias the baseline upward. The corrected spectrum was 
\begin{equation} 
S_{\mathrm{corr}}(f)=S_{\mathrm{avg}}(f)-B(f), 
\end{equation} 
where \(B(f)\) is the fitted baseline. The smoothing parameter was \(\lambda=3\times10^{10}\), the convergence tolerance was \(10^{-5}\), and at most 20 iterations were used. The same baseline procedure was used in the visualization and for the local spectral regions entering the empirical-template fits.

Each peak in the spectrum was unambiguously assigned to the corresponding ion species using two independent approaches. In the first approach, the transport code LISE++\,\cite{tarasov2008_LISE,Tarasov_2016} was utilized to simulate the production, transmission and injection of ions. The results were implemented into the customized code \textsc{RionID} \cite{rionid}. In the second approach, a comprehensive list of ion species was created from the AME2020 database including the isomeric states. In both cases the best match between the measured and reference spectrum was taken as a valid identification. The complication of the data from 410\,MHz cavity is that adjacent harmonics of the revolution frequency overlap and may lead to an erroneous peak assignment. To exclude such possibility, the identification was cross-checked on the spectrum from the 245\,MHz cavity, where only a single harmonics is present. 


\subsection{Peak shape extraction}

\begin{figure}[h!]
\centering
\includegraphics[trim={0 0 0 1.1cm},clip,width=1\textwidth]
{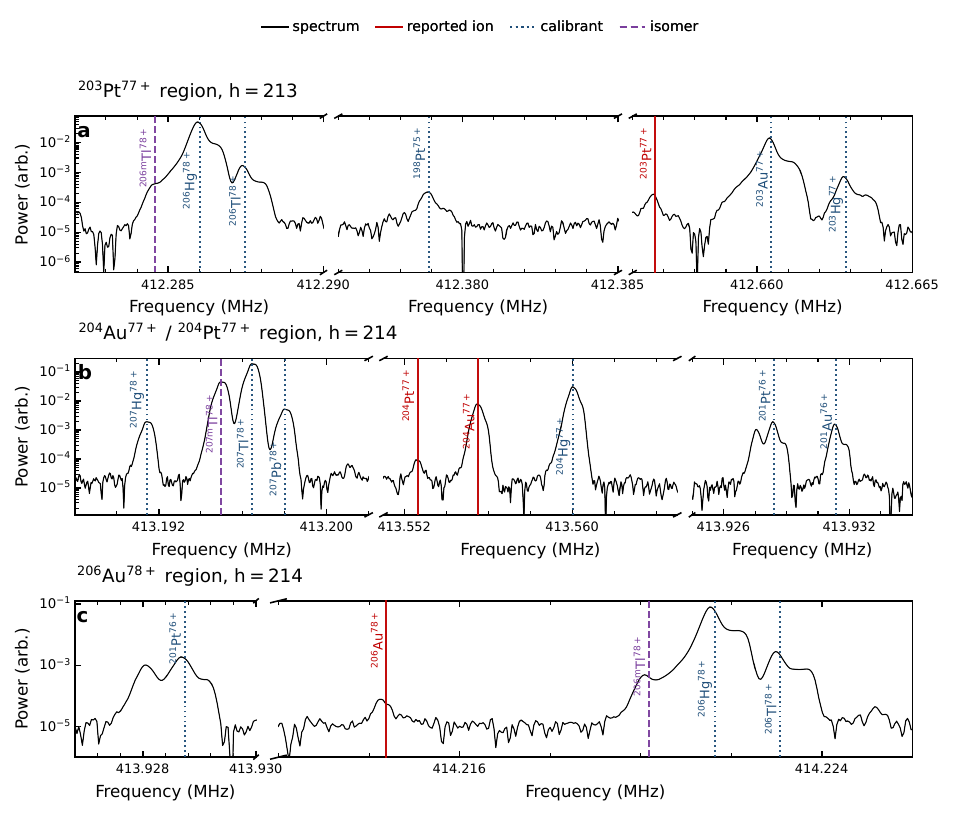}
\caption{\textbf{Extended Data Fig.~1 | Fitting regions for the remaining reported masses.}
\textbf{(a--c)} Regions of the averaged Schottky spectrum used for the local mass determinations of \(^{206}\mathrm{Au}^{78+}\), \(^{203}\mathrm{Pt}^{77+}\), and \(^{204}\mathrm{Au}^{77+}/^{204}\mathrm{Pt}^{77+}\), respectively.
Red markers indicate the ground-state masses reported in the main text; dotted markers indicate nearby reference ions used for local calibration or assignment.
The harmonic number \(h\) is given in each panel, and break marks denote omitted frequency intervals within the same local setting.
}
\label{fig:extended_local_spectra}
\end{figure}

Masses of nuclei of interest were determined through short-range interpolations. The fitting ranges are indicated in Fig.\,\ref{fig:experiment} (b) and zoomed in Fig.\,\ref{fig:experiment} (c) and in Extended Data Fig.\,\ref{fig:extended_local_spectra}.

The Schottky frequency peaks are non-Gaussian and asymmetric, reflecting the remaining non-linearity of $\gamma_t$. In each local region, a nearby high-statistics, uncontaminated reference peak was used to construct an empirical line-shape template \(T(u)\). The template was represented in linear power, normalized to unit area and re-centered to zero centre of gravity. Target regions containing \(K\) components were fitted with

\begin{equation}
M(f)=B(f)+\sum_{k=1}^{K}A_k\frac{1}{s_k}T\!\left(\frac{f-\mu_k}{s_k}\right),
\end{equation}

where \(B(f)\) is a low-order residual background, \(A_k\) is the integrated area of component \(k\), \(s_k\) is a scale factor, and \(\mu_k\) is the fitted centre-of-gravity frequency. The fitted \(\mu_k\) centroid values were used for the mass determination.

Centroid uncertainties were obtained from residual bootstrap resampling and template-bootstrap propagation. Fit-window, background-order, component-number, scale-constraint and template-edge variations were used as systematic checks. For the \(^{205m}\mathrm{Au}^{77+}\)--\(^{205g}\mathrm{Au}^{77+}\) doublet, illustrated in Fig.\,\ref{fig:experiment} (e), the two components were fitted simultaneously with the same empirical template.

\subsection{Local mass calibration}

For a reference ion with mass number \(A\), charge state \(q\), atomic ground-state mass excess \(ME_{\mathrm{gs}}\), charge-state-dependent electron binding energy \(E_{\mathrm{bind}}\), and a possible isomeric excitation energy \(E_{\mathrm{iso}}\), the ionic mass-to-charge ratio was calculated as

\begin{equation}
\left(\frac{m}{q}\right)_{\mathrm{ref}}=\frac{Au+ME_{\mathrm{gs}}+E_{\mathrm{iso}}-qm_e+E_{\mathrm{bind}}}{qu},
\end{equation}
where $u$ stands for the atomic mass unit.
The excitation-energy term was included only for reference ions assigned to known long-lived isomeric states. 

Obtained centroids $\mu$ were transformed to the first harmonic of the revolution frequency, which was done by dividing the measured frequency by the corresponding harmonic number. In each local region, a low-order polynomial function was used to fit the references. The number of references was 4-7 and polynomials up to 3$^{\rm rd}$ order were employed, see more details in Supplementary Materials.


Unknown masses were obtained by evaluating the fitted polynomials and propagating the full coefficient covariance matrix. 
The fitted ionic mass was converted back to the atomic mass.

We employed the standard method to estimate systematic uncertainty of the fitting method\,\cite{Litvinov-2005b}.
Each of the reference masses was considered unknown and re-determined from the remaining reference masses. The literature ($m_{lit}$) and re-determined ($m_{cal}$) masses were compared by building a reduced $\chi^2_{\nu}$-estimator
\begin{equation} 
\chi^2_\nu=\frac{1}{N-p}\sum_{i=1}^N\frac{(m_{lit,i}-m_{cal,i})^2}{\sigma^2_{lit_i}+\sigma^2_{sys}}, 
\end{equation}
where $N$ is the number of reference masses, and $p$ is the number of fitted parameters.
The systematic uncertainty was deduced by forcing the reduced $\chi^2_\nu=1$. Details on all fits are given in Supplementary Materials.


\subsection{Shell-model calculation}

Shell-model calculations were performed in the valence space \(50 \leq Z \leq 82\) and \(82 \leq N \leq 184\). Cross-shell neutron excitations across the \(N=126\) shell gap were not included in the ground-state binding-energy calculations. Two effective interactions were considered. The first one is the Yuan22 interaction introduced in Ref.~\cite{Yuan-2022}, which combines the Kuo--Herling interaction~\cite{Warburton-1991} with the \(V_{\mathrm{MU}}\)LS force~\cite{Liu-2023}. The second one includes the modifications proposed in Ref.~\cite{Liu-2025}, built on Ref.~\cite{Yuan-2022} and incorporating phenomenological adjustments to selected two-body matrix elements introduced in Refs.~\cite{Steer-2011,Yeung-2024}. All calculations were carried out with the KSHELL code~\cite{Shimizu-2019}.

\subsection{Masses of $N=128$ isotones $^{207}$Au and $^{206}$Pt}
The empirical shell-gap indicator $\Delta_{2n}$ requires masses of nuclei beyond those directly measured in this work, specifically the $N=128$ isotones $^{206}$Pt and $^{207}$Au. These masses have not been measured experimentally and were not yet accessible at present facilities. 

The AME extrapolation procedure\,\cite{AME2020} follows the smooth trend of the nuclear mass surface in regions where experimental data are absent, assuming local regularity of the mass surface. Extrapolated values are provided only for a few nuclei in the immediate vicinity of the last known masses, and carry deliberately generous uncertainties to reflect the intrinsic limitations of the approach. These uncertainties are propagated through all finite-difference observables computed here and are reflected in the error bars shown in Fig.\,\ref{fig:deltan} (a, b).
We note that the smooth-surface assumption underlying the AME extrapolation is known to be less reliable in the vicinity of shell closures and/or shape changes, where the mass surface exhibits characteristic discontinuities. 
In this context, our new masses for \(N=125\) and \(N=127\) gold isotopes constrain the local slope on both sides of the closure, whereas for platinum isotopes the \(N=125\) point is available.

The new extrapolations utilized the up-to-date AME network containing all available experimental mass data including our new masses reported here.
The obtained values are listed in Table\,\ref{tab:AME} and will be a part of the forthcoming AME2026 release.
These data were employed to calculate $\Delta_{2n}$ down to iridium ($Z=77$) in Fig.\,\ref{fig:deltan} (a). 
\begin{table}[t] 
\centering 
\caption{\textbf{Extrapolated mass excess values.} Updated AME extrapolations (\(\mathrm{ME}_{\mathrm{AME2026}}\)) and \(\mathrm{ME}_{\mathrm{AME2020}}\)) for the nuclei required to extract two-neutron shell gap, see Fig.\,\ref{fig:deltan}.
The employed up-to-date AME network contains the experimental mass value of $^{201}$Ir, which was adopted here and which will be reported elsewhere.
The symbol \(\#\) denotes extrapolated values.
} \label{tab:AME} \renewcommand{\arraystretch}{1.15} 
\begin{tabular}{lccc} \hline 
Atom & \(N\) & \(\mathrm{ME}_{\mathrm{AME2026}}\) (keV) & \(\mathrm{ME}_{\mathrm{AME2020}}\) (keV) \\ \hline 
\(^{201}\mathrm{Ir}\) & 124 & \(-19617(13)^*\)  & \(-19840(200)\#\) \\ 
\(^{203}\mathrm{Ir}\) & 126 & \(-14370(400)\)\#  & \(-16640(300)\#\)  \\ 
\(^{205}\mathrm{Ir}\) & 128 & \(-5640(500)\)\# & \(-5600(500)\#\) \\ 
\(^{205}\mathrm{Pt}\) & 127 & \(-13040(300)\)\#  & \(-12820(300)\#\) \\ 
\(^{206}\mathrm{Pt}\) & 128 & \(-9550(300)\)\#  & \(-9240(300)\#\)  \\ 
\(^{207}\mathrm{Au}\) & 128 & \(-10850(300)\)\#  & \(-10640(300)\#\)  \\ 
\hline 
\end{tabular} 
~\\
\(^*\) experimental value in the updated AME
\label{tab:table2}
\end{table}

\subsection{Data availability}

The mass values generated in this study are provided in Table~\ref{tab:masses}. The minimum dataset required to reproduce the mass calibration, including processed frequency centroids, reference-ion lists, local calibration inputs, covariance matrices and derived mass-filter tables, will be deposited in a DOI-minting repository before publication. 

\subsection{Code availability}
The analysis used \textsc{RionID}, a publicly available Python package for ion identification, archived under DOI 10.5281/zenodo.8169341 and installable from PyPI. The custom Python routines used to reproduce the figures, tables and reported masses will be deposited in a DOI-minting repository before publication. Code not required for reproducibility of the present results will be documented separately where appropriate.

\subsection{Use of artificial intelligence tools}
Large language models were used during manuscript preparation to assist with wording, structure and copy-editing. The authors reviewed and verified all scientific content, numerical values, interpretations, references and conclusions, and take full responsibility for the final manuscript.

\subsection{Acknowledgments}

The results presented here are based on
the joint experiment G-22-00206 and G-22-00018, which was performed at the ESR
storage ring of the GSI Helmholtzzentrum f{\"u}r
Schwerionenforschung, Darmstadt (Germany) in the context of FAIR Phase-0 research program.

The authors thank the GSI accelerator operations team and the ESR technical staff for excellent support during the experiment. 

Work at Argonne National Laboratory was supported by the U.S. Department of Energy, Office of Science, Office of Nuclear Physics, under Contract Nos. DE-AC02-06CH11357 and DE-SC0023688. X.Z. acknowledges support from the National Natural Science Foundation of China under Grant No. 12305126. H.M. acknowledges support from the German Federal Ministry of Research, Technology and Space (BMFTR) under Grant No. 05P21RDCI2. This work was supported by the Slovenian Research and Innovation Agency (ARIS) under Grant Nos. I0-E005 and P1-0102. D.F.-F. and E.B.M. acknowledge support from the project ``NRW-FAIR'', funded through the programme ``Netzwerke 2021'', an initiative of the Ministry of Culture and Science of the State of North Rhine-Westphalia. C.Y. acknowledges support from the National Natural Science Foundation of China under Grant No. 12475129. This work is supported by the European Research Council (ERC) under the European Union’s Horizon 2020 research and innovation program (ERC-advanced grant NECTAR, Grant No. 884715). 
T.Y. and T.M. acknowledge support from the JSPS KAKENHI Grant Number 23KK0055. M.C., I.D., and C.J.G. acknowledge funding by the Natural Sciences and Engineering Research Council of Canada.

We remember Dr. Markus Steck, who over many years built and refined the experimental capabilities of the ESR storage ring at GSI, enabling a generation of precision experiments including the one reported here. His expertise, dedication and enthusiasm were an inspiration to all who worked with him. 

\subsection{Author contributions}
U.A., H.A., J.B., C.B., J.P.B., R.C., M.C., I.D., D.D., D.F.-F., C.F., O.F., W.G., J.G., M.G., C.J.G., A.G., R.H., P.-M.H., N.J.H., C.J., B.J., K., G.K., F.G.K., W.K., C.K., J.E.L., G.L., H.L., S.L., Y.A.L., Z.L., B.L., H.M., E.B.M., T.M., C.M.N., Z.N., F.C.O., N.P., Z.P., S.S., R.S.S., M.S., T.S., J.V., Q.W., H.W., M.W., K.W., B.W., T.Y., X.X., X.Y., Y.Y., X.Z., and M.Z conducted the experiment. W.K., Y.A.L. and S.S. acted as spokespersons on the experiment. 
R.J.C., S.D., C.F., D.F.-F., J.G., F.K., W.K., G.L., Y.A.L., Z.N., Z.P., R.S.S., M.W., Q.W., P.M.W., X.X., X.Y., Y.Y., and Y.Z. formed the coordination group.
R.C., D.F.-F., X.Y. and Y.Y., conducted the data analysis. 
R.C. and F.G.K. performed updated AME extrapolations. M.L. and C.Y. conducted shell model investigations. 
R.B.C. addressed proton-neutron interaction. R.B.C., D.F.-F., and Y.A.L. drafted the manuscript. All co-authors read and improved the manuscript and contributed to the interpretation of the results. 

\subsection{Competing interests}
The authors declare no competing interests.


\newpage
\section*{Extended Data}

\clearpage

\bibliographystyle{unsrt}

\bibliography{bibliography}

\clearpage

\begin{center}
{\Large\bfseries Supplementary Materials}\\[0.6em]
{\large Precision masses of neutron-rich platinum and gold nuclei reveal enhanced $N=126$ shell strength below doubly-magic $^{208}$Pb}\\[0.6em]
\end{center}

\vspace{1em}

\textbf{The scope of this Supplementary Materials}
is restricted to the quantities reported in the main manuscript, namely the details on the determination the ground-state masses of \nuc{203}{Pt}, \nuc{204}{Pt}, \nuc{204}{Au}, \nuc{205g}{Au}, and \nuc{206}{Au}, together with the resolved isomeric state \nuc{205m}{Au}. Provided are the calibration, uncertainty treatment, and consistency checks that support the reported mass surface across the $N=126$ shell closure.

\section*{Supplementary Note 1: local mass calibration and uncertainty treatment}

Masses were determined from local Schottky frequency intervals rather than from a single global calibration, see Fig.\,2 in the manuscript. In each interval the measured revolution-frequency coordinate $x$ was mapped to ionic mass-to-charge ratio through a polynomial function
\begin{equation}
  \left(\frac{m}{q}\right)(x)=\sum_{k=0}^{p-1} a_k z^k,
  \qquad
  z=\frac{x-x_0}{x_{\mathrm{scale}}}.
\end{equation}
Here $x_0$ and $x_{\mathrm{scale}}$ are numerical centering and scaling constants chosen separately for each local setting. The polynomial order was selected based on the result of the re-determination of calibration masses routine, see Methods, and on the frequency distribution of calibrant peaks. Unknown masses were then obtained by evaluating the fitted polynomial at the corresponding empirical-template centroid of the Schottky peak, see Methods, and propagating the centroid uncertainty and the full calibration covariance matrix.

Ionic mass-to-charge ratios were converted to neutral-atom mass excesses by adding the corresponding electron masses and applying the charge-state-dependent electron-binding correction, see Methods. This conversion was applied consistently to all ions, and the reported mass excess values (\(ME\)) correspond to neutral atoms throughout. 

The uncertainties quoted for each reported state are \sigfitsys. The $\sigma_{\rm fit}$ contains the uncertainties of the peak-centroid determination, the propagated local-calibration covariance, and the contribution due to calibrant-mass errors. The method to estimate $\sigma_{\rm sys}$ is described in Methods. It is based on the re-determination of every reference mass assuming it is unknown.  

\begin{table}[H]
\centering
\caption{Details for the reported ground-state masses.
The listed \(f_{\rm rev}\) values correspond to the real revolution frequency of the ions in the ESR (first harmonic).
The mass excesses are given for neutral atoms. 
$\Delta ME$ is the difference to the value extrapolated in the Atomic Mass Evaluation (AME2020). 
}\label{tab:reported_masses}
\small
\begin{tabular}{llrrrrrr}
\toprule
State  & $q$ & $f_{\rm rev}$ (MHz) & $\sigma_f{_{\rm rev}}$ (Hz) & $ME$ (keV) & $\Delta ME$ (keV) & \sigfitsys{} (keV) \\
\midrule
\nuc{203}{Pt}     & 77+ & 1.937356 & 0.04319 & -19741.88 & -231.88 & 37.07 \\
\nuc{204}{Pt}  & 77+ & 1.932489 & 0.03521 & -18023.27 & -403.27 & 6.92 \\
\nuc{204}{Au}   & 77+ & 1.932502 & 0.00497 & -20618.36 & -228.36 & 1.75 \\
\nuc{205}{Au}   & 77+ & 1.927650 & 0.04488 & -19034.32 & -464.32 & 10.18 \\
\nuc{206}{Au}    & 78+ & 1.935581 & 0.03402 & -14322.94 & -132.94 & 20.65 \\

\bottomrule
\end{tabular}
\end{table}


\section*{Supplementary Note 2: local calibration settings}

\begin{table}[H]
\centering
\caption{Local calibration settings for the reported masses. The independent determination of \nuc{205g}{Au} and \nuc{205m}{Au} masses for $q=78+$ charge state is provided as an additional consistency check; the reported values are from the $q=77+$ setting. Listed are the employed polynomial fir degree (Degree), number of free parameters ($p$), number of calibration masses ($N_{cal}$), and the reduced $\chi^2_\nu$ of the systematic error estimation routine, see Methods.}
\label{tab:calibration_settings}
\small
\begin{tabular}{llrrrrl}
\toprule
Setting & Reported state(s) & Degree & $p$ & $N_{\mathrm{cal}}$ & $\sigma_{\mathrm{sys}}$ (keV) & 
$\chi^2_\nu$
\\
\midrule
\nuc{205}{Au}, $q=77+$ & \nuc{205m}{Au}, \nuc{205g}{Au} & 3 & 4 & 7 & 0 & \(\chi^2_\nu=0.3\) \\
\nuc{204}{Au}--\nuc{204}{Pt} & \nuc{204}{Au}, \nuc{204}{Pt} & 2 & 3 & 6 & 0 & \(\chi^2_\nu=0.3\) \\
\nuc{206}{Au} & \nuc{206}{Au} & 1 & 2 & 4 & 18.89 & $\chi^2_{\nu}=1.5$ \\
\nuc{203}{Pt} & \nuc{203}{Pt} & 1 & 2 & 5 & 36.06 & $\chi^2_{\nu}=25.8$ \\
\nuc{205}{Au}, $q=78+$ & consistency check only & 2 & 3 & 6 & 85.71 & $\chi^2_{\nu}=27.5$ \\
\bottomrule
\end{tabular}
\end{table}

The \nuc{204}{Au}--\nuc{204}{Pt} setting has a special local geometry. The reported \(A=204\) peaks lie very close to the \qion{204}{Hg}{77} reference peak, whereas the remaining reference ions form two more distant groups. The \(A=204\) masses are therefore best described as locally anchored interpolations constrained primarily by \qion{204}{Hg}{77} and validated by the surrounding reference network. The dominance of the local anchor is quantified in Supplementary Table~\ref{tab:dominant_calibrants}.

\begin{longtable}{llrrrr}
\caption{Re-determined masses of reference ions in the local settings. The nearby \qion{204}{Hg}{77} anchor was not removed in the \nuc{204}{Au}--\nuc{204}{Pt} validation because omitting it would turn the test into a long-range extrapolation.
Listed are the revolution frequencies of the reference ions ($f_{rev}$), deviation to the tabulated value from AME2020 ($\Delta ME_{\mathrm{LOO}}$), the combined uncertainty ($\sigma_{\mathrm{LOO+AME2020}}$), and the empirical systematic uncertainty
\(\sigma_{\mathrm{sys}}\) assigned to the corresponding local setting.
}\label{tab:loo_validation}\\
\toprule
Setting & Reference ion & $f_{\rm rev}$ (MHz) & $\Delta ME_{\mathrm{LOO}}$ (keV) & $\sigma_{\mathrm{LOO+AME2020}}$ (keV) & $\sigma_{\mathrm{sys}}$ \\
\midrule
\endfirsthead
\caption[]{Re-determined masses of reference ions in the local settings. Continued.}\\
\toprule
Setting & Calibrant & $f_{\rm rev}$ (MHz) & $\Delta ME_{\mathrm{LOO}}$ (keV) & $\sigma_{\mathrm{LOO+AME2020}}$ (keV) & $\sigma_{\mathrm{sys}}$ \\
\midrule
\endhead
\midrule
\multicolumn{6}{r}{Continued on next page}\\
\endfoot
\bottomrule
\endlastfoot
\nuc{205}{Au}, $q=77+$ & \qion{208}{Tl}{78} & 1.926025 & -13.21 & 22.71 &  0.00 \\
\nuc{205}{Au}, $q=77+$ & \qion{208}{Pb}{78} & 1.926050 & +12.76 & 22.00 &  0.00 \\
\nuc{205}{Au}, $q=77+$ & \qion{205}{Hg}{77} & 1.927666 & +5.05 & 11.29  &  0.00 \\
\nuc{205}{Au}, $q=77+$ & \qion{205}{Tl}{77} & 1.927674 & -5.33 & 11.35  &  0.00 \\
\nuc{205}{Au}, $q=77+$ & \qion{202}{Pt}{76} & 1.929326 & +19.95 & 26.18 &  0.00 \\
\nuc{205}{Au}, $q=77+$ & \qion{202}{Au}{76} & 1.929335 & +4.57 & 23.93  &  0.00 \\
\nuc{205}{Au}, $q=77+$ & \qion{202}{Hg}{76} & 1.929351 & -12.52 & 18.64 &  0.00 \\
\addlinespace
\nuc{204}{Au}--\nuc{204}{Pt} & \qion{207}{Hg}{78} & 1.930801 & -15.04 & 38.03 &  0.00 \\
\nuc{204}{Au}--\nuc{204}{Pt} & \qion{207}{Tl}{78} & 1.930825 & +2.14 & 8.85   &  0.00 \\
\nuc{204}{Au}--\nuc{204}{Pt} & \qion{207}{Pb}{78} & 1.930832 & -2.15 & 11.31  &  0.00 \\
\nuc{204}{Au}--\nuc{204}{Pt} & \qion{201}{Pt}{76} & 1.934245 & -24.08 & 50.92 &  0.00 \\
\nuc{204}{Au}--\nuc{204}{Pt} & \qion{201}{Au}{76} & 1.934259 & +24.27 & 51.33 &  0.00 \\
\addlinespace
\nuc{206}{Au} & \qion{201}{Pt}{76} & 1.934245 & -51.86 & 50.10   & 35.75 \\
\nuc{206}{Au} & \qion{201}{Au}{76} & 1.934259 & +50.70 & 49.73   & 35.75 \\
\nuc{206}{Au} & \qion{206}{Hg}{78} & 1.935615 & +22.86 & 21.36   & 35.75 \\
\nuc{206}{Au} & \qion{206}{Tl}{78} & 1.935622 & -23.06 & 21.47   & 35.75  \\
\addlinespace
\nuc{203}{Pt} & \qion{206m}{Tl}{78} & 1.935608 & -14.76 & 13.29 & 36.06 \\
\nuc{203}{Pt} & \qion{206}{Hg}{78} & 1.935615 & +24.30 & 21.19  & 36.06  \\
\nuc{203}{Pt} & \qion{206}{Tl}{78} & 1.935622 & +2.58 & 11.90   & 36.06  \\
\nuc{203}{Pt} & \qion{203}{Au}{77} & 1.937373 & +40.61 & 4.68   & 36.06 \\
\nuc{203}{Pt} & \qion{203}{Hg}{77} & 1.937384 & -40.88 & 4.71   & 36.06 \\
\addlinespace
\nuc{205}{Au}, $q=78+$ & \qion{203}{Au}{77} & 1.937373 & +38.93 & 4.71  & 85.71\\
\nuc{205}{Au}, $q=78+$ & \qion{203}{Hg}{77} & 1.937384 & -39.28 & 4.68  & 85.71 \\
\nuc{205}{Au}, $q=78+$ & \qion{200}{Pt}{76} & 1.939186 & +52.86 & 21.79 & 85.71  \\
\nuc{205}{Au}, $q=78+$ & \qion{205}{Hg}{78} & 1.940426 & -45.97 & 22.42 & 85.71  \\
\nuc{205}{Au}, $q=78+$ & \qion{202}{Pt}{77} & 1.942241 & +102.90 & 34.16& 85.71  \\
\nuc{205}{Au}, $q=78+$ & \qion{202}{Au}{77} & 1.942250 & -78.00 & 34.18 & 85.71 \\
\end{longtable}


\section*{Supplementary Note 3: Flow of information}

\begin{table}[H]
\centering
\caption{Principal calibrant influences on the reported masses. The variance fractions quantify the relative weight of each listed calibrant in the final mass determination. Only the dominant contributors are given.}
\label{tab:dominant_calibrants}
\small
\begin{tabular}{lll}
\toprule
Reported mass & Dominant calibrant(s) & Variance fraction(s) \\
\midrule
\qion{205m}{Au}{77} & \qion{205}{Hg}{77}; \qion{205}{Tl}{77}; \qion{208}{Tl}{78} & 0.61147; 0.19890; 0.17200 \\
\qion{205g}{Au}{77} & \qion{205}{Hg}{77}; \qion{205}{Tl}{77}; \qion{208}{Tl}{78} & 0.65196; 0.21909; 0.11731 \\
\qion{204}{Pt}{77} & \qion{204}{Hg}{77} & 0.99929 \\
\qion{204}{Au}{77} & \qion{204}{Hg}{77} & 0.99973 \\
\qion{206}{Au}{78} & \qion{206}{Tl}{78}; \qion{206}{Hg}{78} & 0.91944; 0.08022 \\
\qion{203}{Pt}{77} & \qion{203}{Hg}{77}; \qion{203}{Au}{77} & 0.53526; 0.46385 \\
\bottomrule
\end{tabular}
\end{table}
The high precision of the reported mass values of \nuc{204}{Au} and \nuc{204}{Pt} reflects their
proximity to \qion{204}{Hg}{77}. The results are obtained via local interpolations
anchored by the nearest reference ion, while the surrounding reference network
tests the consistency of the local calibration curve.

\section*{Supplementary Note 4: \nuc{205}{Au} doublet}

The reported \nuc{205}{Au} values are obtained from the $q=77+$ setting. In this setting, the resolved doublet gives
\begin{equation}
  ME(\nuc{205m}{Au})=-18104.64(9.52)_{\mathrm{fit+sys}}~\mathrm{keV},
\end{equation}
\begin{equation}
  ME(\nuc{205g}{Au})=-19034.32(10.18)_{\mathrm{fit+sys}}~\mathrm{keV}.
\end{equation}
The resulting excitation energy of the resolved isomeric partner is
\begin{equation}
  E_x(\nuc{205m}{Au})=929.68(13.94)~\mathrm{keV},
\end{equation}
where the uncertainty is an independent quadrature of the two \sigfitsys{} values. Correlations in the local calibration would reduce this uncertainty, so this estimate is deliberately conservative.

\begin{table}[H]
\centering
\caption{Independent consistency check for the \nuc{205}{Au} doublet. The $q=78+$ setting has been used only to test the consistency with $q=77+$ setting. 
}\label{tab:q78_consistency}
\small
\begin{tabular}{llrrrr}
\toprule
State & Charge state & $f_{\rm rev}$ (MHz) & $ME$ (keV) & \sigfitsys{} (keV) & Difference from $q=77+$ (keV) \\
\midrule
\nuc{205m}{Au} & $77+$ & 1.927645 & -18104.64 & 9.52 & -- \\
\nuc{205m}{Au} & $78+$ & 1.940404 & -18065.93 & 89.06 & +38.71(89.57) \\
\nuc{205g}{Au} & $77+$ & 1.927650 & -19034.32 & 10.18 & -- \\
\nuc{205g}{Au} & $78+$ & 1.940408 & -18938.95 & 89.29 & +95.37(89.87) \\
\bottomrule
\end{tabular}
\end{table}

\section*{Supplementary Note 5: Figure 1\,b data}

\begin{sidewaystable}[p] 
\centering 
\footnotesize 
\setlength{\tabcolsep}{4.0pt} 
\renewcommand{\arraystretch}{1.15} 
\begin{threeparttable} 
\caption{\textbf{Mass-excess values used in Fig.~1.} Numerical inputs for the \(N=126\) isotonic mass-excess comparison. All values are neutral-atom mass excesses in MeV. The AME2020 column is retained separately to show the extrapolated values replaced by the new measurements. The plotted residuals in Fig.~1 are obtained by subtracting the predictions by Duflo---Duker (DZ) mass model, \(\Delta\mathrm{ME}=\mathrm{ME}-\mathrm{ME}_{\mathrm{DZ}}\).} \label{tab:fig1a_mass_excess_inputs} \begin{tabular}{lccrrrrlrrrrr} \toprule Nuclide & \(Z\) & \(N\) & \(\mathrm{ME}_{\mathrm{AME20}}\) & \(\sigma_{\mathrm{AME20}}\) & \(\mathrm{ME}_{\mathrm{exp}}\) & \(\sigma_{\mathrm{exp}}\) & Source & FRDM12 & WS4+RBF & DZ & BSk21 & Yuan22 SM \\ \midrule \(^{202}\mathrm{Os}\) & 76 & 126 & \(-12.530\) & 0.400 & \(-12.530\) & 0.400 & AME2020\# & \(-13.780\) & \(-12.658\) & \(-13.610\) & \(-13.580\) & \(-12.842\) \\ \(^{203}\mathrm{Ir}\) & 77 & 126 & \(-14.370\) & 0.400 & \(-14.370\) & 0.400 & AME2020\# & \(-15.260\) & \(-14.408\) & \(-15.230\) & \(-15.440\) & \(-14.839\) \\ \(^{204}\mathrm{Pt}\) & 78 & 126 & \(-17.620\) & 0.200 & \(-18.023\) & 0.007 & This work & \(-18.330\) & \(-17.807\) & \(-18.400\) & \(-18.220\) & \(-17.803\) \\ \(^{205}\mathrm{Au}\) & 79 & 126 & \(-18.570\) & 0.200 & \(-19.034\) & 0.010 & This work & \(-18.880\) & \(-18.514\) & \(-19.120\) & \(-19.060\) & \(-19.043\) \\ \(^{206}\mathrm{Hg}\) & 80 & 126 & \(-20.945728\) & 0.020 & \(-20.945728\) & 0.020 & AME2020 & \(-20.880\) & \(-21.063\) & \(-21.360\) & \(-21.110\) & \(-20.927\) \\ \(^{207}\mathrm{Tl}\) & 81 & 126 & \(-21.034436\) & 0.005 & \(-21.034436\) & 0.005 & AME2020 & \(-20.410\) & \(-20.818\) & \(-21.060\) & \(-21.210\) & \(-21.072\) \\ \(^{208}\mathrm{Pb}\) & 82 & 126 & \(-21.748519\) & 0.011 & \(-21.748519\) & 0.011 & AME2020 & \(-20.920\) & \(-21.810\) & \(-22.290\) & \(-22.060\) & \(-21.786\) \\ \bottomrule \end{tabular} \begin{tablenotes} \footnotesize \item \(\#\) denotes an AME2020 value extrapolated from systematic trends. \item FRDM12 denotes the Finite-Range Droplet Model; WS4+RBF denotes the Weizs\"acker--Skyrme mass model with radial-basis-function correction; DZ denotes the Duflo--Zuker mass model used as the baseline for the residuals plot; BSk21 denotes the Hartree-Fock-Bogoliubov model with Brussels--Skyrme-v.21 functional; Yuan22 SM denotes the shell-model calculation. 
\end{tablenotes} 
\end{threeparttable} 
\end{sidewaystable}

\clearpage

\section*{Supplementary Note 6: Figure 3 data}
The numerical values entering Fig.~3 (a) are listed separately for the shifted shell gap \(\Delta_{2n}(Z,N=124)\) in Supplementary
Table~\ref{tab:fig3_delta2n_minus_data} and the conventional shell gap \(\Delta_{2n}(Z,N=126)\) in Supplementary
Table~\ref{tab:fig3_delta2n_plus_data}. The one-neutron shell gap used in Figs.~3 (b,c) are listed in Supplementary
Table~\ref{tab:fig3_delta1n_data}.

\begin{sidewaystable}[p]
\centering
\scriptsize
\setlength{\tabcolsep}{3.0pt}
\renewcommand{\arraystretch}{1.13}
\begin{threeparttable}
\caption{\textbf{Shifted two-neutron shell gap used in Fig.~3 (a).}
Values of \(\Delta_{2n}(Z,124)=S_{2n}(Z,124)-S_{2n}(Z,126)\). 
The column ``Mass combination'' lists the mass-excess coefficients entering the finite difference,
\(\Delta_{2n}(Z,124)=ME(Z,122)-2ME(Z,124)+ME(Z,126)\).
Rows with missing mass components are not plotted.}
\label{tab:fig3_delta2n_minus_data}
\begin{tabular}{lcclrrll}
\toprule
Variant  & \(Z\) & Element & Nuclide & Value (MeV) & Unc. (MeV) & Mass combination & Source category \\
\midrule
AME2020\#  & 76 & Os & \(^{202}\mathrm{Os}\) & 0.970 & 0.748 & \(^{198}\mathrm{Os}:+1;\ ^{200}\mathrm{Os}:-2;\ ^{202}\mathrm{Os}:+1\) & AME2020\# \\
AME2020\#  & 77 & Ir & \(^{203}\mathrm{Ir}\) & 0.911 & 0.567 & \(^{199}\mathrm{Ir}:+1;\ ^{201}\mathrm{Ir}:-2;\ ^{203}\mathrm{Ir}:+1\) & AME2020\# \\
NewExp+AME2026\# & 77 & Ir & \(^{203}\mathrm{Ir}\) & 0.465 & 0.403 & \(^{199}\mathrm{Ir}:+1;\ ^{201}\mathrm{Ir}:-2;\ ^{203}\mathrm{Ir}:+1\) & NewExp+AME2026\# \\
AME2020\#  & 78 & Pt & \(^{204}\mathrm{Pt}\) & 1.165 & 0.207 & \(^{200}\mathrm{Pt}:+1;\ ^{202}\mathrm{Pt}:-2;\ ^{204}\mathrm{Pt}:+1\) & AME2020\# \\
NewExp  & 78 & Pt & \(^{204}\mathrm{Pt}\) & 0.762 & 0.055 & \(^{200}\mathrm{Pt}:+1;\ ^{202}\mathrm{Pt}:-2;\ ^{204}\mathrm{Pt}:+1\) & NewExp+AME2020 \\
AME2020\#  & 79 & Au & \(^{205}\mathrm{Au}\) & 1.316 & 0.200 & \(^{201}\mathrm{Au}:+1;\ ^{203}\mathrm{Au}:-2;\ ^{205}\mathrm{Au}:+1\) & AME2020\# \\
NewExp  & 79 & Au & \(^{205}\mathrm{Au}\) & 0.852 & 0.012 & \(^{201}\mathrm{Au}:+1;\ ^{203}\mathrm{Au}:-2;\ ^{205}\mathrm{Au}:+1\) & NewExp+AME2020 \\
AME2020 & 80 & Hg & \(^{206}\mathrm{Hg}\) & 1.089 & 0.020 & \(^{202}\mathrm{Hg}:+1;\ ^{204}\mathrm{Hg}:-2;\ ^{206}\mathrm{Hg}:+1\) & AME2020 \\
AME2020 & 81 & Tl & \(^{207}\mathrm{Tl}\) & 0.846 & 0.006 & \(^{203}\mathrm{Tl}:+1;\ ^{205}\mathrm{Tl}:-2;\ ^{207}\mathrm{Tl}:+1\) & AME2020 \\
AME2020 & 82 & Pb & \(^{208}\mathrm{Pb}\) & 0.713 & 0.003 & \(^{204}\mathrm{Pb}:+1;\ ^{206}\mathrm{Pb}:-2;\ ^{208}\mathrm{Pb}:+1\) & AME2020 \\
AME2020 & 83 & Bi & \(^{209}\mathrm{Bi}\) & 0.785 & 0.007 & \(^{205}\mathrm{Bi}:+1;\ ^{207}\mathrm{Bi}:-2;\ ^{209}\mathrm{Bi}:+1\) & AME2020 \\
AME2020 & 84 & Po & \(^{210}\mathrm{Po}\) & 0.797 & 0.005 & \(^{206}\mathrm{Po}:+1;\ ^{208}\mathrm{Po}:-2;\ ^{210}\mathrm{Po}:+1\) & AME2020 \\
AME2020 & 85 & At & \(^{211}\mathrm{At}\) & 0.893 & 0.016 & \(^{207}\mathrm{At}:+1;\ ^{209}\mathrm{At}:-2;\ ^{211}\mathrm{At}:+1\) & AME2020 \\
AME2020 & 86 & Rn & \(^{212}\mathrm{Rn}\) & 0.895 & 0.014 & \(^{208}\mathrm{Rn}:+1;\ ^{210}\mathrm{Rn}:-2;\ ^{212}\mathrm{Rn}:+1\) & AME2020 \\
AME2020 & 87 & Fr & \(^{213}\mathrm{Fr}\) & 0.944 & 0.027 & \(^{209}\mathrm{Fr}:+1;\ ^{211}\mathrm{Fr}:-2;\ ^{213}\mathrm{Fr}:+1\) & AME2020 \\
AME2020 & 88 & Ra & \(^{214}\mathrm{Ra}\) & 0.933 & 0.023 & \(^{210}\mathrm{Ra}:+1;\ ^{212}\mathrm{Ra}:-2;\ ^{214}\mathrm{Ra}:+1\) & AME2020 \\
AME2020 & 89 & Ac & \(^{215}\mathrm{Ac}\) & 0.892 & 0.060 & \(^{211}\mathrm{Ac}:+1;\ ^{213}\mathrm{Ac}:-2;\ ^{215}\mathrm{Ac}:+1\) & AME2020 \\
AME2020 & 90 & Th & \(^{216}\mathrm{Th}\) & 1.020 & 0.026 & \(^{212}\mathrm{Th}:+1;\ ^{214}\mathrm{Th}:-2;\ ^{216}\mathrm{Th}:+1\) & AME2020 \\
AME2020 & 91 & Pa & \(^{217}\mathrm{Pa}\) & 1.100 & 0.175 & \(^{213}\mathrm{Pa}:+1;\ ^{215}\mathrm{Pa}:-2;\ ^{217}\mathrm{Pa}:+1\) & AME2020 \\
\bottomrule
\end{tabular}
\begin{tablenotes}
\footnotesize
\item The \(^{218}\mathrm{U}\) \(\Delta_{2n}(N=124)\) value is not plotted because the \(N=122\) component is missing.
\item Filled-star \(\Delta_{2n}(Z,N=124)\) values for Pt and Au do not require \(N=128\) masses.
\item The symbol \(\#\) denotes a value that contains at least one extrapolated AME input.
\end{tablenotes}
\end{threeparttable}
\end{sidewaystable}


\begin{sidewaystable}[p]
\centering
\scriptsize
\setlength{\tabcolsep}{3.0pt}
\renewcommand{\arraystretch}{1.13}
\begin{threeparttable}
\caption{\textbf{Two-neutron shell-gap used in Fig.~3 (a).}
Values of \(\Delta_{2n}(Z,126)=S_{2n}(Z,126)-S_{2n}(Z,128)\). 
The column ``Mass combination'' lists the mass-excess coefficients entering the finite difference,
\(\Delta_{2n}(Z,126)=ME(Z,124)-2ME(Z,126)+ME(Z,128)\).
Rows with missing mass components are not plotted.}
\label{tab:fig3_delta2n_plus_data}
\begin{tabular}{lcclrrll}
\toprule
Variant  & \(Z\) & Element & Nuclide & Value & Unc. & Mass combination & Source category \\
\midrule
AME2020\# & 77 & Ir & \(^{203}\mathrm{Ir}\) & 3.300 & 0.964 & \(^{201}\mathrm{Ir}:+1;\ ^{203}\mathrm{Ir}:-2;\ ^{205}\mathrm{Ir}:+1\) & AME2020\# \\
NewExp+AME2026\# & 77 & Ir & \(^{203}\mathrm{Ir}\) & 3.483 & 0.943 & \(^{201}\mathrm{Ir}:+1;\ ^{203}\mathrm{Ir}:-2;\ ^{205}\mathrm{Ir}:+1\) & NewExp+AME2026\# \\
AME2020\# & 78 & Pt & \(^{204}\mathrm{Pt}\) & 3.308 & 0.501 & \(^{202}\mathrm{Pt}:+1;\ ^{204}\mathrm{Pt}:-2;\ ^{206}\mathrm{Pt}:+1\) & AME2020\# \\
NewExp+AME2026\#  & 78 & Pt & \(^{204}\mathrm{Pt}\) & 3.804 & 0.302 & \(^{202}\mathrm{Pt}:+1;\ ^{204}\mathrm{Pt}:-2;\ ^{206}\mathrm{Pt}:+1\) & NewExp+AME2026\# \\
AME2020\#  & 79 & Au & \(^{205}\mathrm{Au}\) & 3.357 & 0.500 & \(^{203}\mathrm{Au}:+1;\ ^{205}\mathrm{Au}:-2;\ ^{207}\mathrm{Au}:+1\) & AME2020\# \\
NewExp+AME2026\# & 79 & Au & \(^{205}\mathrm{Au}\) & 4.075 & 0.301 & \(^{203}\mathrm{Au}:+1;\ ^{205}\mathrm{Au}:-2;\ ^{207}\mathrm{Au}:+1\) & NewExp+AME2026\# \\
AME2020  & 80 & Hg & \(^{206}\mathrm{Hg}\) & 3.936 & 0.051 & \(^{204}\mathrm{Hg}:+1;\ ^{206}\mathrm{Hg}:-2;\ ^{208}\mathrm{Hg}:+1\) & AME2020 \\
AME2020  & 81 & Tl & \(^{207}\mathrm{Tl}\) & 4.603 & 0.013 & \(^{205}\mathrm{Tl}:+1;\ ^{207}\mathrm{Tl}:-2;\ ^{209}\mathrm{Tl}:+1\) & AME2020 \\
AME2020  & 82 & Pb & \(^{208}\mathrm{Pb}\) & 4.983 & 0.003 & \(^{206}\mathrm{Pb}:+1;\ ^{208}\mathrm{Pb}:-2;\ ^{210}\mathrm{Pb}:+1\) & AME2020 \\
AME2020  & 83 & Bi & \(^{209}\mathrm{Bi}\) & 4.604 & 0.007 & \(^{207}\mathrm{Bi}:+1;\ ^{209}\mathrm{Bi}:-2;\ ^{211}\mathrm{Bi}:+1\) & AME2020 \\
AME2020  & 84 & Po & \(^{210}\mathrm{Po}\) & 4.068 & 0.003 & \(^{208}\mathrm{Po}:+1;\ ^{210}\mathrm{Po}:-2;\ ^{212}\mathrm{Po}:+1\) & AME2020 \\
AME2020  & 85 & At & \(^{211}\mathrm{At}\) & 3.831 & 0.009 & \(^{209}\mathrm{At}:+1;\ ^{211}\mathrm{At}:-2;\ ^{213}\mathrm{At}:+1\) & AME2020 \\
AME2020  & 86 & Rn & \(^{212}\mathrm{Rn}\) & 3.394 & 0.012 & \(^{210}\mathrm{Rn}:+1;\ ^{212}\mathrm{Rn}:-2;\ ^{214}\mathrm{Rn}:+1\) & AME2020 \\
AME2020  & 87 & Fr & \(^{213}\mathrm{Fr}\) & 3.286 & 0.017 & \(^{211}\mathrm{Fr}:+1;\ ^{213}\mathrm{Fr}:-2;\ ^{215}\mathrm{Fr}:+1\) & AME2020 \\
AME2020  & 88 & Ra & \(^{214}\mathrm{Ra}\) & 2.907 & 0.017 & \(^{212}\mathrm{Ra}:+1;\ ^{214}\mathrm{Ra}:-2;\ ^{216}\mathrm{Ra}:+1\) & AME2020 \\
AME2020  & 89 & Ac & \(^{215}\mathrm{Ac}\) & 2.782 & 0.030 & \(^{213}\mathrm{Ac}:+1;\ ^{215}\mathrm{Ac}:-2;\ ^{217}\mathrm{Ac}:+1\) & AME2020 \\
AME2020  & 90 & Th & \(^{216}\mathrm{Th}\) & 2.465 & 0.027 & \(^{214}\mathrm{Th}:+1;\ ^{216}\mathrm{Th}:-2;\ ^{218}\mathrm{Th}:+1\) & AME2020 \\
AME2020  & 91 & Pa & \(^{217}\mathrm{Pa}\) & 2.278 & 0.111 & \(^{215}\mathrm{Pa}:+1;\ ^{217}\mathrm{Pa}:-2;\ ^{219}\mathrm{Pa}:+1\) & AME2020 \\
AME2020\#  & 92 & U & \(^{218}\mathrm{U}\) & 2.290 & 0.108 & \(^{216}\mathrm{U}:+1;\ ^{218}\mathrm{U}:-2;\ ^{220}\mathrm{U}:+1\) & AME2020\# \\
\bottomrule
\end{tabular}
\begin{tablenotes}
\footnotesize
\item The \(^{202}\mathrm{Os}\) \(\Delta_{2n}(N=126)\) value is not plotted because \(^{204}\mathrm{Os}\) is not available in the adopted surface.
\item For Pt and Au, the open-star \(\Delta_{2n}(Z,N=126)\) values contain extrapolated \(N=128\) neighbours, \(^{206}\mathrm{Pt}\) and \(^{207}\mathrm{Au}\), respectively.
\item The symbol \(\#\) denotes a value that contains at least one extrapolated AME input.
\end{tablenotes}
\end{threeparttable}
\end{sidewaystable}

\begin{sidewaystable}[p]
\centering
\scriptsize
\setlength{\tabcolsep}{3.2pt}
\renewcommand{\arraystretch}{1.13}
\begin{threeparttable}
\caption{\textbf{One-neutron shell gap used in Fig.~3 (b,c).}
Values of \(\Delta_{1n}(Z,N)=S_n(Z,N)-S_n(Z,N+1)\) for Pt and Au. 
The mass combination column lists the neutral-atom mass-excess coefficients entering the finite difference.}
\label{tab:fig3_delta1n_data}
\begin{tabular}{llcclrrll}
\toprule
Variant & Element & \(Z\) & \(N\) & Nuclide & Value (MeV) & Unc. (MeV) & Mass combination & Source \\
\midrule

AME2020 & Pt & 78 & 120 & \(^{198}\mathrm{Pt}\) & 2.000 & 0.005 & \(^{197}\mathrm{Pt}:+1;\ ^{198}\mathrm{Pt}:-2;\ ^{199}\mathrm{Pt}:+1\) & AME2020 \\
AME2020 & Pt & 78 & 122 & \(^{200}\mathrm{Pt}\) & 2.069 & 0.064 & \(^{199}\mathrm{Pt}:+1;\ ^{200}\mathrm{Pt}:-2;\ ^{201}\mathrm{Pt}:+1\) & AME2020 \\
AME2020\# & Pt & 78 & 124 & \(^{202}\mathrm{Pt}\) & 2.134 & 0.212 & \(^{201}\mathrm{Pt}:+1;\ ^{202}\mathrm{Pt}:-2;\ ^{203}\mathrm{Pt}:+1\) & AME2020\# \\
NewExp & Pt & 78 & 124 & \(^{202}\mathrm{Pt}\) & 1.902 & 0.072 & \(^{201}\mathrm{Pt}:+1;\ ^{202}\mathrm{Pt}:-2;\ ^{203}\mathrm{Pt}:+1\) & NewExp+AME2020 \\
AME2020\# & Pt & 78 & 126 & \(^{204}\mathrm{Pt}\) & 2.910 & 0.539 & \(^{203}\mathrm{Pt}:+1;\ ^{204}\mathrm{Pt}:-2;\ ^{205}\mathrm{Pt}:+1\) & AME2020\# \\
NewExp+AME2020\# & Pt & 78 & 126 & \(^{204}\mathrm{Pt}\) & 3.484 & 0.301 & \(^{203}\mathrm{Pt}:+1;\ ^{204}\mathrm{Pt}:-2;\ ^{205}\mathrm{Pt}:+1\) & NewExp+AME2020\# \\
AME2020\# & Pt & 78 & 128 & \(^{206}\mathrm{Pt}\) & 1.520 & 0.781 & \(^{205}\mathrm{Pt}:+1;\ ^{206}\mathrm{Pt}:-2;\ ^{207}\mathrm{Pt}:+1\) & AME2020\# \\
AME2026\# & Pt & 78 & 128 & \(^{206}\mathrm{Pt}\) & 2.140 & 0.781 & \(^{205}\mathrm{Pt}:+1;\ ^{206}\mathrm{Pt}:-2;\ ^{207}\mathrm{Pt}:+1\) & AME2026\#+AME2020\# \\

\addlinespace
AME2020 & Au & 79 & 120 & \(^{199}\mathrm{Au}\) & 1.367 & 0.027 & \(^{198}\mathrm{Au}:+1;\ ^{199}\mathrm{Au}:-2;\ ^{200}\mathrm{Au}:+1\) & AME2020 \\
AME2020 & Au & 79 & 122 & \(^{201}\mathrm{Au}\) & 1.208 & 0.036 & \(^{200}\mathrm{Au}:+1;\ ^{201}\mathrm{Au}:-2;\ ^{202}\mathrm{Au}:+1\) & AME2020 \\
AME2020\# & Au & 79 & 124 & \(^{203}\mathrm{Au}\) & 1.544 & 0.201 & \(^{202}\mathrm{Au}:+1;\ ^{203}\mathrm{Au}:-2;\ ^{204}\mathrm{Au}:+1\) & AME2020\# \\
NewExp & Au & 79 & 124 & \(^{203}\mathrm{Au}\) & 1.316 & 0.026 & \(^{202}\mathrm{Au}:+1;\ ^{203}\mathrm{Au}:-2;\ ^{204}\mathrm{Au}:+1\) & NewExp+AME2020 \\
AME2020\# & Au & 79 & 126 & \(^{205}\mathrm{Au}\) & 2.560 & 0.539 & \(^{204}\mathrm{Au}:+1;\ ^{205}\mathrm{Au}:-2;\ ^{206}\mathrm{Au}:+1\) & AME2020\# \\
NewExp & Au & 79 & 126 & \(^{205}\mathrm{Au}\) & 3.128 & 0.024 & \(^{204}\mathrm{Au}:+1;\ ^{205}\mathrm{Au}:-2;\ ^{206}\mathrm{Au}:+1\) & NewExp \\
AME2020\# & Au & 79 & 128 & \(^{207}\mathrm{Au}\) & 1.180 & 0.735 & \(^{206}\mathrm{Au}:+1;\ ^{207}\mathrm{Au}:-2;\ ^{208}\mathrm{Au}:+1\) & AME2020\# \\
NewExp+AME2026\# & Au & 79 & 128 & \(^{207}\mathrm{Au}\) & 1.468 & 0.671 & \(^{206}\mathrm{Au}:+1;\ ^{207}\mathrm{Au}:-2;\ ^{208}\mathrm{Au}:+1\) & NewExp+AME2026\# \\
AME2020\# & Au & 79 & 130 & \(^{209}\mathrm{Au}\) & 1.230 & 0.943 & \(^{208}\mathrm{Au}:+1;\ ^{209}\mathrm{Au}:-2;\ ^{210}\mathrm{Au}:+1\) & AME2020\# \\
\bottomrule
\end{tabular}
\begin{tablenotes}
\footnotesize
\item The \(^{208}\mathrm{Pt}\) \(\Delta_{1n}\) point is not plotted because the \(N=131\) component is missing.
\item For Au at \(N=126\), the updated \(\Delta_{1n}\) value is determined entirely from the three consecutive masses \(^{204}\mathrm{Au}\), \(^{205}\mathrm{Au}\), and \(^{206}\mathrm{Au}\) measured in this work. This point is therefore plotted as a filled red star in Fig.~3 (c).
\item The Pt \(N=126\) updated \(\Delta_{1n}\) value contains the extrapolated \(^{205}\mathrm{Pt}\) mass and is therefore plotted as an open red star in Fig.~3 (b).
\end{tablenotes}
\end{threeparttable}
\end{sidewaystable}

\begin{table}[H]
\centering
\small
\caption{\textbf{Shell-model two-neutron shell gap used in Fig.~3 (a).}
Values of \(\Delta_{2n}(N=126)\) and \(\Delta_{2n}(N=124)\) calculated from the Yuan22 shell-model binding energies.}
\label{tab:fig3_yuan_delta2n}
\begin{tabular}{cclrr}
\toprule
\(Z\) & Element & \(N\) & \(\Delta_{2n}(N=126)\) (MeV) & \(\Delta_{2n}(N=124)\) (MeV) \\
\midrule
76 & Os & 126 & 3.206 & 0.599 \\
77 & Ir & 126 & 3.508 & 0.665 \\
78 & Pt & 126 & 3.522 & 0.696 \\
79 & Au & 126 & 3.860 & 0.803 \\
80 & Hg & 126 & 3.958 & 0.962 \\
81 & Tl & 126 & 4.574 & 0.785 \\
82 & Pb & 126 & 5.015 & 0.652 \\
\bottomrule
\end{tabular}
\end{table}

\begin{table}[H]
\centering
\small
\caption{\textbf{Shell-model one-neutron shell gap used in Fig.~3 (b,c).}
Values of \(\Delta_{1n}\) calculated from the Yuan22 shell-model binding energies. }
\label{tab:fig3_yuan_delta1n}
\begin{tabular}{cclrr}
\toprule
\(Z\) & Element & \(N\) & Nuclide & \(\Delta_{1n}\) (MeV)\\
\midrule
78 & Pt & 122 & \(^{200}\mathrm{Pt}\) & 2.042 \\
78 & Pt & 124 & \(^{202}\mathrm{Pt}\) & 1.698 \\
78 & Pt & 126 & \(^{204}\mathrm{Pt}\) & 2.915 \\
78 & Pt & 128 & \(^{206}\mathrm{Pt}\) & 1.409 \\
79 & Au & 120 & \(^{199}\mathrm{Au}\) & 1.567 \\
79 & Au & 122 & \(^{201}\mathrm{Au}\) & 1.466 \\
79 & Au & 124 & \(^{203}\mathrm{Au}\) & 1.344 \\
79 & Au & 126 & \(^{205}\mathrm{Au}\) & 2.853 \\
79 & Au & 128 & \(^{207}\mathrm{Au}\) & 1.208 \\
\bottomrule
\end{tabular}
\end{table}

\section*{Supplementary Note 7: Figure 4 data}

\begin{sidewaystable}[p]
\centering
\scriptsize
\setlength{\tabcolsep}{4.0pt}
\renewcommand{\arraystretch}{1.12}
\begin{threeparttable}

\caption{\textbf{Average proton--neutron interaction strengths used in Fig.~4.}
The table lists the updated \(\delta V_{pn}\) values obtained from the mass surface used in this work. All \(\delta V_{pn}\) values and uncertainties are given in keV. The labels \(ee\) and \(oe\) denote the corresponding even--even and odd-\(Z\), even-\(N\) mass filters defined in Methods.}
\label{tab:fig4_dvpn_inputs}

\begin{tabular}{ccclccrl}
\toprule
\(Z\) &
Element &
\(N\) &
Nuclide &
Filter &
\(\delta V_{pn}\) (keV) &
\(\sigma\) (keV) &
Mass inputs entering the updated value \\
\midrule
78 & Pt & 120 & \(^{198}\mathrm{Pt}\) & \(ee\) &
354.4 & 10.0 & AME2020 \\

79 & Au & 120 & \(^{199}\mathrm{Au}\) & \(oe\) &
347.3 & 1.1 & AME2020 \\
79 & Au & 122 & \(^{201}\mathrm{Au}\) & \(oe\) &
305.9 & 10.2 & AME2020 \\
79 & Au & 124 & \(^{203}\mathrm{Au}\) & \(oe\) &
324.9 & 16.3 & AME2020 \\
79 & Au & 126 & \(^{205}\mathrm{Au}\) & \(oe\) &
279.8 & 14.5 &
AME2020 + \(^{204}\mathrm{Pt}\), \(^{205}\mathrm{Au}\) from this work \\

80 & Hg & 120 & \(^{200}\mathrm{Hg}\) & \(ee\) &
322.4 & 0.6 & AME2020 \\
80 & Hg & 122 & \(^{202}\mathrm{Hg}\) & \(ee\) &
286.7 & 5.1 & AME2020 \\
80 & Hg & 124 & \(^{204}\mathrm{Hg}\) & \(ee\) &
313.0 & 8.1 & AME2020 \\
80 & Hg & 126 & \(^{206}\mathrm{Hg}\) & \(ee\) &
231.2 & 8.5 &
AME2020 + \(^{204}\mathrm{Pt}\) from this work \\

81 & Tl & 120 & \(^{201}\mathrm{Tl}\) & \(oe\) &
286.2 & 15.7 & AME2020 \\
81 & Tl & 122 & \(^{203}\mathrm{Tl}\) & \(oe\) &
369.2 & 7.1 & AME2020 \\
81 & Tl & 124 & \(^{205}\mathrm{Tl}\) & \(oe\) &
357.3 & 1.0 & AME2020 \\
81 & Tl & 126 & \(^{207}\mathrm{Tl}\) & \(oe\) &
479.0 & 10.6 & AME2020 \\

82 & Pb & 120 & \(^{202}\mathrm{Pb}\) & \(ee\) &
285.2 & 2.7 & AME2020 \\
82 & Pb & 122 & \(^{204}\mathrm{Pb}\) & \(ee\) &
331.8 & 1.0 & AME2020 \\
82 & Pb & 124 & \(^{206}\mathrm{Pb}\) & \(ee\) &
332.7 & 0.5 & AME2020 \\
82 & Pb & 126 & \(^{208}\mathrm{Pb}\) & \(ee\) &
426.9 & 5.1 & AME2020 \\
\bottomrule
\end{tabular}

\begin{tablenotes}
\footnotesize
\item The updated \(^{205}\mathrm{Au}\) value uses the new \(^{205}\mathrm{Au}\) and \(^{204}\mathrm{Pt}\) masses. The updated \(^{206}\mathrm{Hg}\) value uses the new \(^{204}\mathrm{Pt}\) mass. 
\item Values for Os are not listed because the corresponding \(ee\) filters require neighbouring \(Z=74\) masses that are not available in the adopted mass surface.
\item Uncertainties were obtained by Gaussian propagation of the input mass uncertainties. Correlations between masses are not included unless explicitly stated in the calibration covariance.
\item AME2020 denotes the evaluated atomic-mass surface used for the remaining input masses. Where AME2020 extrapolated values enter a derived observable in other tables, they are indicated by the standard \(\#\) notation.
\end{tablenotes}

\end{threeparttable}
\end{sidewaystable}

The tabulated values in Supplementary Notes 5--7 are the numerical source data
for Figs.~1, 3 and~4. Machine-readable versions will be deposited together with
the processed calibration inputs in the data repository as described in the Data
Availability statement.

\end{document}